\documentclass[
superscriptaddress,
 amsmath,amssymb,
 aps,
prper,
twocolumn,
]{revtex4-1}

\usepackage{parskip}
\usepackage{enumitem}
\usepackage{xcolor}
\usepackage{graphicx}% Include figure files
\usepackage{dcolumn}% Align table columns on decimal point
\usepackage{bm}% bold math

\usepackage{url}

\usepackage{xr}% cross reference to other files (supplemental.aux - get aux file from compiling supplemental.tex, then click "Other logs & files", click aux, upload in Overleaf and change name to supplemental.aux  )
\begin{document}

\title{The impact of content sequencing on quantitative problem-solving: A case study in Electromagnetism using an online learning platform}

\author{Benjamin J. Dringoli}
\affiliation{Department of Physics, McGill University, 3600 rue University, Montr\'{e}al, Qu\'{e}bec H3A2T8, Canada}
\author{Ksenia Kolosova}
\affiliation{Department of Physics, McGill University, 3600 rue University, Montr\'{e}al, Qu\'{e}bec H3A2T8, Canada}
\author{Thomas J. Rademaker}
\affiliation{Department of Physics, McGill University, 3600 rue University, Montr\'{e}al, Qu\'{e}bec H3A2T8, Canada}
\author{Juliann Wray}
\affiliation{Department of Physics, McGill University, 3600 rue University, Montr\'{e}al, Qu\'{e}bec H3A2T8, Canada}
\author{Jeremie Choquette}
\affiliation{Department of Physics, McGill University, 3600 rue University, Montr\'{e}al, Qu\'{e}bec H3A2T8, Canada}
\affiliation{Dawson College, 3040 Sherbrooke St W, Montr\'{e}al, Qu\'{e}bec H3Z 1A4, Canada}
\author{Michael Hilke}
\email{hilke@physics.mcgill.ca}
\affiliation{Department of Physics, McGill University, 3600 rue University, Montr\'{e}al, Qu\'{e}bec H3A2T8, Canada}

\date{\today}% It is always \today, today, but any date may be explicitly specified

\begin{abstract}
We investigate the impact of content sequencing on quantitative problem-solving in a first-year university Electromagnetism and Optics course using an online learning platform. With the custom-built McGill Learning Platform (McLEAP), we test students' quantitative problem-solving ability as a function of the sequence in which the students are presented new educational material. New educational material was divided into conceptual, theoretical, and example-based content. We find that with over 1500 students in two years of study, students presented with conceptual content first perform better on our assessment than those presented with theoretical content first, and interpret this finding through the lens of the constructivist theory of learning. Additionally, we find that instructors’ preferences for content sequencing differ strongly from that of students. We discuss some of the challenges associated with online learning tools and conclude that McLEAP is easily generalizable to other academic levels and courses, and could even be adapted to measure conceptual understanding.

\end{abstract}

\maketitle

%\tableofcontents

\section{Introduction}
\label{sec:intro}

\parindent=15pt
\parskip=5pt

For physics instructors, a fundamental question underlies the evaluation of lectures, lesson plans, and other learning materials: ``Is there a way to present educational content that results in optimal learning for students?'' While this question is simple, its investigation incorporates psychology, neuroscience, education research, and subject-specific considerations \cite{Thambyah_2011,Brewe_2018,Sternberg_1995,Davis_2009,Jones_1987}. Moreover, to understand which teaching material is best for student learning, we must understand both how content can be categorized in the subject of interest and how variations between strategies can be measured reliably from student data.

The investigation of how different teaching approaches impact student learning is an active area in  Science, Technology, Engineering and Mathematics (STEM) education research \cite{brown2012current}, due in part to the increased focus on proficiency in science and math related subjects in the past decades \cite{NRC}, especially at the university level. In addition, investigations of this kind are a unique opportunity to probe subtle differences between the preferences of new and experienced learners, such as populations of students and professors, to better facilitate knowledge transfer between the two populations.

While teaching an introductory Electromagnetism and Optics course for first-year university STEM students, we began to investigate these questions around which characteristics of educational content impact its teaching effectiveness. Many proposed theories exist for how students learn best \cite{anderson_2001,mcleod_2017,gardner_2011}, but not all take into account the effects of content types or sequencing. We note that within physics pedagogy, content type categorization is implicit in many educational materials. Content type categories commonly encountered in physics textbooks and courses include: theories, concepts, examples, historical contexts, and laboratory experiments, among others. In physics textbook design, the way in which these content types are presented inherits the writer's personal pedagogical goals and teaching style: certain content types may be more prominent, content types may be presented in a certain order, and some content types may be left out. Textbooks and corresponding teaching manuals are then used as templates for course and laboratory design by instructors, and as learning material for students. Since these choices are present in all physics instruction, we became interested in how factors like content type, content order, student preferences, and instructor preferences affect student learning and performance.

In this case study, we probe how the sequence in which three common physics content types were presented contributed to students' learning outcomes in an online learning module. We randomly assign sequences of concept, theory, and example content to students and evaluate the impact on student problem-solving using various metrics of normalized gain. To accomplish this, we introduce the McGill Learning Platform (McLEAP) tool, which offers a flexible online space for investigating pedagogical questions such as these, and discuss analysis methods and results from two years of student study using this tool.

\section{Background}
\label{sec:background}

\subsection{Physics Content Types}

We choose to divide educational content into three types, which form the basis of our study into teaching material: \textbf{Concepts}, \textbf{Examples}, and \textbf{Theories}. We consider these categories to be independent for this investigation, while noting it is possible for content to contain aspects of multiple types, and that in certain frameworks Concept and Theory may both be categorized under conceptual learning \cite{krathwohl2002}. This is clearly not the only way to partition teaching material, but we are interested if this specific partitioning (which we call ``ECT categorization'') can be shown to affect student learning or borrows strategies from any prior work on improving teaching effectiveness in physics. Below we will outline our original definitions of these content types, as well as some similar descriptions found in previous studies around Physics Education Research (PER).`

Concepts, in the context of our study, consist of general information and visual explanations allowing for broad connections to scientific principles. They relate students' preconceived notions of physics and lived experiences to more structured, illustrative, and physically intuitive information, often accompanied by visual aids. We view and define the Concept category similarly to that of Van Heuvelen \cite{OCS} as well as Gautreau and Novemsky \cite{GautreauNovemsky}: as a qualitative overview of the physics at hand, using words, sketches, diagrams and/or graphs. 

Theories represent the rigorous mechanics behind the phenomenon of interest, often accompanied or assisted by mathematical equations or descriptions. This categorization of the formal mathematics and definitions of physics content is similar to the approach of more 'mathematical-based' teaching \cite{OCS}. Theory contrasts Concept in that it is a concrete and formal description of the phenomenon; its role is to emphasize the mathematical expression which exactly describes the physics \cite{Karam2014,Chassy2019}, and its basis is less related to the pre-conceived notions or real-world analogies that a student may encounter with the Concept content type.

Examples, in our interpretation, serve as road maps which show the correct setup and solution method to a commonly encountered type of problem. They consist of fully completed solutions to quantitative problems similar to those found on quizzes and exams. This definition aligns with the 'worked example-based' content in Renkl \cite{renkl2014,renkl2016} and the 'worked example principle' used in Booth and co-workers \cite{Booth}. This content type specifically trains learners in thought processes required for solving physics problems, but not necessarily why certain concepts or theories are used. %BD: Is this last sentence okay? I feel like it adds a good solid contrast to the other types.

Each of these content types are shown to be effective parts of instruction for introductory physics content. Further, some studies show that exposure to more than one type of these content categories will greatly benefit learning content \cite{OCS, 5E}. This further motivates our interest in finding sequences of content types that maximize student learning. %BD: not sure if this paragraph is necessary given the same info is presented in the first paragraph of the next section.
%TR: I think it is a good intro into the sequencing section

\subsection{Sequencing}

We define a content sequence as an ordered list of the three types of physics instructional content, such as \{Concept, Theory, Example\} (CTE) or  \{Example, Concept, Theory\} (ECT). There are therefore six unique content sequences that can be made with our three content type categories. Multiple studies \cite{OCS,5E} have found evidence to suggest that students' learning outcomes can be impacted by exposure to several content types which describe and represent the same physical situation. We use these sequences to evaluate the effects of being exposed to content types at different stages of the learning process. This is accomplished by analyzing the impact of sequencing on students' problem-solving performance.

Content sequencing forms the basis of the influential BSCS 5E instructional model, a planning tool for instructors proposed by Bybee et al. \cite{bybee2006}. Tanner explicitly studied the optimal order for sequencing elements of instruction to maximize student learning in an introductory biology course setting through the lens of the 5E model \cite{5E}. The 5E model was previously reviewed within earlier research on the learning cycle, which specifically investigated the effects of changing the sequence \cite{RennerAbrahamBirnie1988,MarekCavallo}. Their work indicated reduced effectiveness of the 5E model when the sequence was changed, which further bolsters the hierarchical nature of this model. Similarly to these studies, we explore the effectiveness of six different permutations of our content types (Concept, Theory, and Example) as sequences and whether certain sequences allow students to perform better than others within the scope of our introductory Electromagnetism and Optics course. 

The ordering of physics content is also examined in the Overview Case Study Physics by Van Heuvelen \cite{OCS}, reviewed by Gautreau and Novemsky \cite{GautreauNovemsky}, where the main approach is: only after the basic physics concepts are understood do students get involved in mathematical solutions to problems. From there, an emphasis is placed on multiple-representation problem-solving (using pictoral, physical, and mathematical methods as mentioned in the previous section), and practicing different permutations of these multiple representations. The preliminary results of this study, which compared a conventional approach with the Van Heuvelen approach, showed that using their approach allows for higher gains in student's qualitative understanding, their problem-solving ability, and their ability to form and access a knowledge hierarchy compared to a more conventional approach in teaching introductory level physics. Here, students taught in the conventional way had a $6\%$ gain, while the gain of students taught in the OCS way was $20\%$, showing significant improvements when exposure was had to a hierarchy of content types \cite{OCS}. 

Content sequencing is also related to the framework of scaffolding, or setting up learning such that students have sufficient prior knowledge to learn complex topics and solve problems \cite{lindstrom2011teaching}. Podolefsky et al. have conducted studies in Electromagnetism classes that found that scaffolding by teaching students new topics using analogies improves student learning outcomes \cite{podolefsky2007analogical, podolefsky2007analogical2}. This type of analogical scaffolding relates to our definition of Concept, as it links student knowledge to information that they are already comfortable with, such as scenarios from the natural world, and uses diagrammatic representations to cement new ideas. Results from this literature thus motivate the use of content sequences that begin by providing students with a strong conceptual base. 

In looking into the the effectiveness of these different permutations, we ultimately examine whether there exists a optimal hierarchy of content (with our choice of partitioning) for students to learn from. The previously mentioned studies have a focus on a concept-first approach, while others are more oriented towards an example-based learning approach. \cite{Booth} concludes that it is best to replace some practice problems with a study of the worked out solution to a given problem. Additionally, the Theory of Example-based learning and suggests that example-based instruction provides students with an initial advantage in the learning process which leads to superior skill acquisition \cite{renkl2014,renkl2016}. These differences may be due to different teaching goals, or different interpretations of how the learning process occurs.

\subsection{Theories of Learning}
\label{Theories}

To understand how these different types of educational content could affect student learning, one must also explore how students learning occurs. As such, this has been a important topic in the field of education research for many years \cite{Shavelson_1972,posner_strike_hewson_gertzog_1982,Redish_1994,Kaser_2017}. We will review two previously proposed theories of student learning which helped inform our understanding in this study: the Revised Bloom's Taxonomy and the Constructivist Theory of Learning.

%as a sectioned pyramid (Fig. \ref{fig:blooms-MI}, left)
Bloom's Taxonomy (BT) posits that there exists a hierarchy of objectives, that certain teaching material can be more effective in teaching, and that there may exist an optimal order for the presentation of content that maximizes student understanding. The revised version modernizes its vocabulary to emphasize the dynamic nature of learning and the different cognitive processes required \cite{Amer2006}. Even with these revisions, 'Bloom’s Revised Taxonomy' still can be visualized, like many hierarchical systems, as a sectioned pyramid. This interpretation ranks the desired outcomes of teaching, with more ‘surface-level’ or basic understanding types (which we place under conceptual understanding) making up the foundation. These then support the ‘higher-level’ learning outcomes which require more expertise and familiarity \cite{Anderson_2005,Murphy_2007}. The higher-level outcomes, in turn, may be similar to those expected from theoretical treatments in later lessons or advanced courses, including advanced textbooks as discussed in Section \ref{Texts}.

% and avoiding misinterpretations (Fig. \ref{fig:blooms-MI}, right)
This system is contrasted by another framework that seeks to address the structure of learning with a less rigid approach than BT: the Constructivist Theory (CT). Instead of prescribing certain objectives as more fundamental than others, the CT places priority on a correct global interpretation and integration of information rather than an absolute sequence in which teaching material is presented \cite{Mevarech1999}. This framework is also related to Generative Theories of learning (GT) \cite{Fiorella2016}, which places significance on prior knowledge and how easily new information can be integrated with the current picture in the learner's mind. While the CT and GT also support a sequence of learning like BT, they do not present a rigid structure, which can support students of various backgrounds and with different amounts of prior knowledge. Instead, they focus on properly integrating information and avoiding misinterpretations. The study of misconceptions has its own body of literature, but we will limit our discussion to the Constructivist framework in this work, while noting the future research into misconception models is of interest to us for future investigations using the McLEAP tool.

It is also interesting to consider the intersections of rigid and flexible theories of learning, as it is possible to imagine a situation where the CT works to direct students to the correct understanding through a certain type of understanding, given a student population with similar initial views on the topic. This creates a quasi-sequenced learning strategy, but enabled by being flexible to student needs. This is the interpretation we take forward, trying to find where students need the most support in introductory university courses by considering different sequences of instruction. %BD: this is a rough paragraph to try to connect this theories discussion to our further work, but I'm not sure if it really flows into the textbook section correctly yet.

\subsection{Textbook Approaches}
\label{Texts}

Finally, to see how these various interpretations of content sequencing and theories of learning have been integrated into content presentation approaches taken by the wider physics education community, we will explore some popular textbooks aimed at teaching Electromagnetism to university students and how they would fit into our content partitioning and sequencing scheme.

A general physics textbook used in first-year courses, such as `Physics' by Giancoli \cite{giancoli}, or the concept-focused `Conceptual Physics' by Hewitt \cite{hewitt2002conceptual}, commonly starts with a broad explanation of the topics being introduced, setting the stage before introducing mathematical forms, and finishes each section with a set of practice problems and some completed problems as a guide. With our partitioning, this represents a concept-first approach, and may take inspiration from the CT, using clear anecdotes to make sure a proper understanding is built before presenting more complete descriptions.

This sequence is not used by all physics textbooks, though. The older classic `Electromagnetic Theory' by Stratton \cite{stratton2007electromagnetic}, a widely referenced undergraduate text, delves right into the mathematical description of Maxwell's equations with little context or examples. This is also the strategy taken by the popular graduate-level text by Jackson \cite{jackson1999classical}, 'Classical Electrodynamics', as it assumes some prior knowledge of the subject. By assuming that the students have the necessary background information to understand theory, this text could represent the middle tiers of BT, representing more complex learning modes (analysis, evaluation) that cannot be accomplished without that fundamental foundation. 

Finally, the popular textbook `Introductory Electromagnetics' by the Popovic father and daughter pair \cite{popovic2000introductory} offers another perspective on how Electromagnetism can be taught. Their textbook builds discussion through worked out examples before introducing a more formal approach, catering to its largely application-centric engineering student audience. Looking at these common resources, we see that there are many divergent ways to approach teaching physics, depending on the teaching goals of the instructor and experience level of the students. As our study focuses on first-year university students, though, we use the evaluation of the Giancoli and Hewitt texts and the previous literature on teaching students new to physics to initially propose the sequence CTE as most likely to be beneficial.

\subsection{Preferences}
\label{Preferences}

Finally, we are interested to see if there are any trends in student preference between the various sequence options. Previous research has shown that students can approach physics problems in varying ways according to their personality \cite{alfathy}, and that students respond favorably when the instruction aligns with a CT-inspired framework \cite{cannon}. %While we were unable to find any previous studies on student preferences for different educational content presentations, % I think this study mentioned is similar enough that we don't need to preface it with this first part of the sentence maybe. -JW
%TR:  Sounds good
A similar study partitioned quiz questions into different representations (verbal, mathematical, graphical, or pictorial) \cite{kohl}. While they found that the students' opinions on which type of question they preferred were self-consistent, they often preferred types of problems that they did not perform the best on. Additionally, student performance was highly dependent on their prior knowledge and the specific format of the question, regardless of the topic or approach being evaluated. We are interested in seeing whether the students in our study know which teaching content type will be the most useful for them during learning, as well as how that preference changes with experience in solving physics problems (for example: undergraduate vs. graduate students). Additionally, we are interested if the choice of content order has a measurable effect even with variation in major parameters like question format and physics topic. This is accomplished by a content type identification and preference survey of our local physics department and students enrolled in the introductory physics classes.

\section{Methods}
\label{sec:methods}

\subsection{Survey}
\label{sec:survey}

To probe the degree of agreement with our educational content classification, and to compare content sequence preferences of the introductory course students to those more familiar with physics topics, we issued an online survey within the local STEM community. Participants were asked to self-identify their field/major (e.g. physics, chemistry) and role (e.g. undergraduate student, graduate student, faculty).

This survey allowed for differences in classification and preference to be quantified across multiple disciplines as well as level of expertise with physics. The participants within the physics department were contacted via departmental email, and further information about other disciplines was gained through distributing the survey to the undergraduates who had taken the introductory Electricity and Magnetism course, concentrated across the sciences. Additional responses were solicited through social media posts, attracting some responses from outside the McGill community, but a majority ($>95$\%) came from within the faculties of science and engineering of McGill University.

111 total participants consented to sharing their survey results for research purposes; we first asked them to assess the accuracy of our content groupings, and then the effectiveness of each content sample in teaching. For the first 9 content samples, participants were asked to label content type between Example, Concept, and Theory. For the following 9 content samples, participants were asked to rate the accuracy of a given label, and the content’s effectiveness in teaching the topic contained, both on a scale of 1 (not accurate/effective) to 5 (very accurate/effective). Three of these samples were mislabeled, serving as a control for response bias. The participants were lastly asked if they have a preferred content type between Example, Concept, and Theory, if there is an order for presenting educational material they think is best between all permutations of Examples, Concept, and Theory, and for any comments and concerns about the content organization in general.

Results of the survey are provided in Figs. S\ref{suppfig:accuracy-labelling} and S\ref{suppfig:rating-content}. We find that content of the Example content type is correctly labeled by 91\% of the survey respondents (Fig. S\ref{suppfig:accuracy-labelling}). Concept is labeled correctly 62\%, similar to Theory (60\%), but the mislabels are distributed between Example and Theory, while for the Theory content type, the mislabels concentrate on Concept. These results are not surprising given that Concept and Theory concern conceptual learning, while Example concerns procedural learning \cite{krathwohl2002}. Moreover, survey respondents were not provided with our definition of Concept, Example and Theory. Note that if the content types were fully orthogonal, our results on sequencing of the three content types would be even stronger. Inevitably, results regress to the mean when contents that are assigned Theory can be partly considered Concept too.

The rating of the content type at describing the content is higher for the correctly labeled content types than it is for the falsely labeled content types (Fig. S\ref{suppfig:rating-content}, left panel), while the effectiveness of the content at teaching the topic is very similar for each of the content types, with Concept being slightly higher than the others (Fig. S\ref{suppfig:rating-content}, right panel).

To verify the reliability of the survey results, we compute the inter-rater reliability index Cronbach's alpha $\rho_T$ on the accuracy of the content label at describing the content for subsets of questions \cite{cronbach1951}. Computed on both the correctly and incorrectly labeled questions we find, $\rho_T = 0.66$, on the subset of correctly labeled content we find $\rho_T = 0.79$ and on the subset of incorrectly labeled content we find $\rho_T = 0.43$. The absolute value of Cronbach's alpha is difficult to interpret given its dependence on the number of topics rated, leading to wide margins that are considered acceptable in the science education research \cite{taber2018}, although $\rho_T = 0.79$ for the subset of correctly labeled questions should be considered satisfactory. We can also compare the value between two sets of questions following \cite{mun2015}. Cronbach's value increases when removing the incorrectly labeled questions (control questions), demonstrating that the survey respondents are reliable raters of correctly labeled questions, meaning they provide similar ratings of the content label.

\subsection{Experimental Assignment: McLEAP}
\label{sec:assignment}

\begin{figure*}[p]
\includegraphics[width=0.8\textwidth]{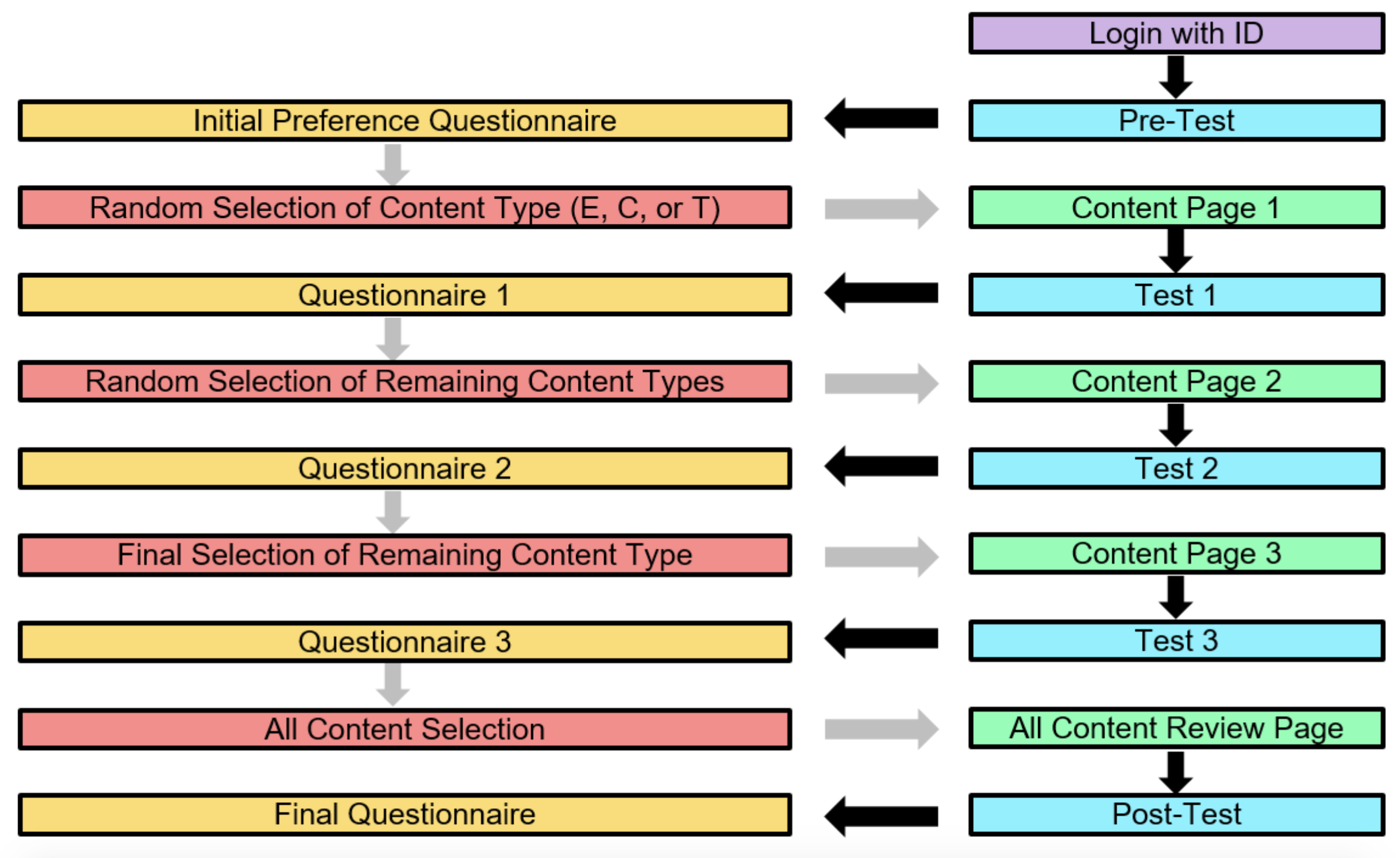}
\caption{Flow chart diagram of McLEAP assignment implementation, showing the mechanical steps the students took to progress and what was presented. The list of questions asked during each of the questionnaires is presented in Fig. S\ref{suppfig:mcleap-interface}. }
\label{fig:mcleap-flowchart}
\end{figure*}

\begin{figure*}[p]
\includegraphics[width=\textwidth]{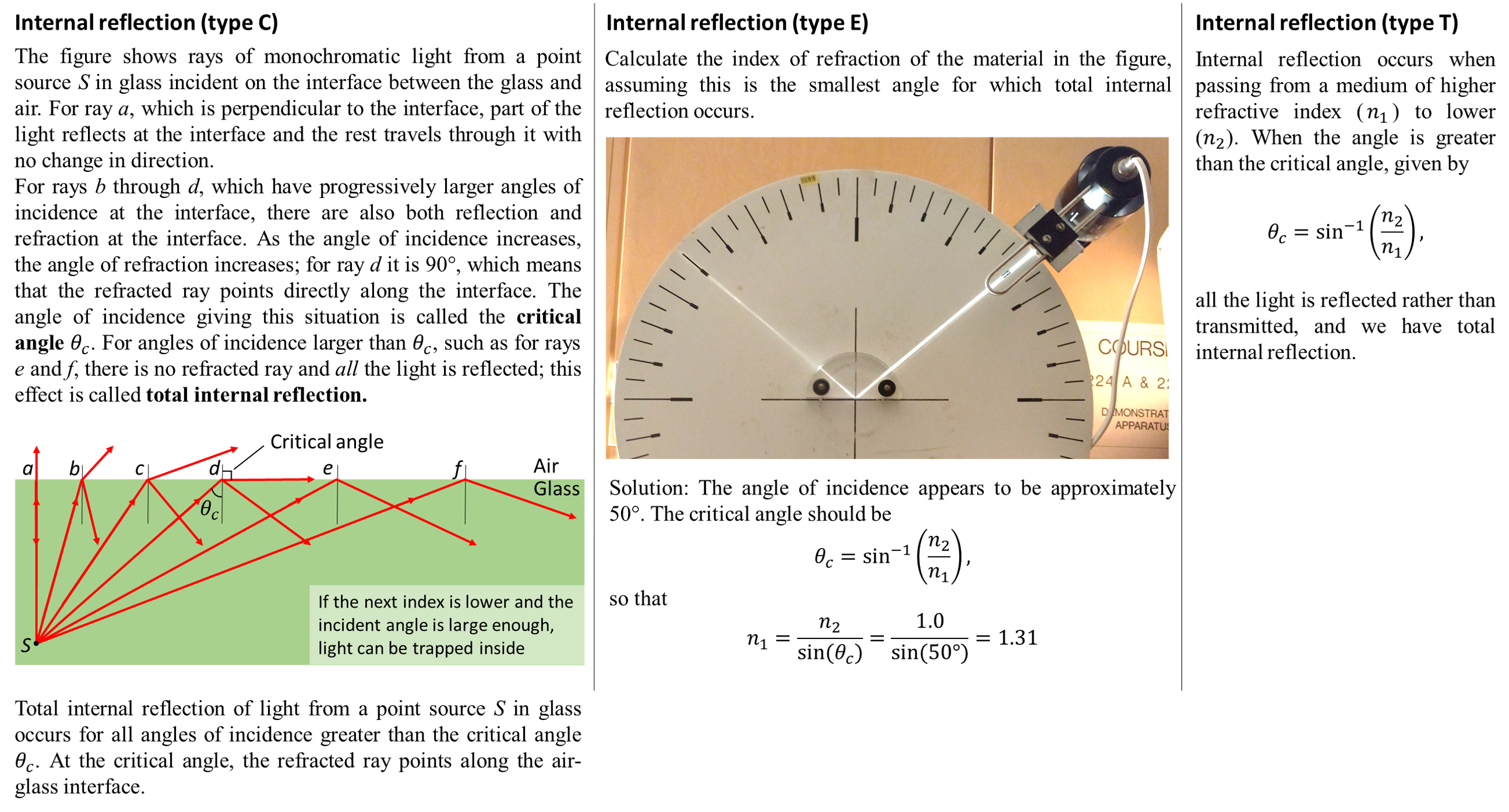}
\caption{Reproduced McLEAP content pages illustrating the three different content types used for classifying educational material into groups and in the assignment study. Left represents a typical concept (the content was adapted from \cite{halliday_2011}), middle a typical example, and right a typical theory.}
\label{fig:illustration-content-type}
\end{figure*}

Once the goals of the study were identified and our content partitioning was chosen, an assignment was designed to test student learning directly and collect relevant information. This was done through an online system where the students fill out five evaluations: a pre-test, three intermediate tests (quiz 1 to 3), and a post-test. In Fig. S\ref{suppfig:quiz} we show a screen capture of the McLEAP tool to illustrate the design and implementation of an intermediate test. 

The full methodological format is shown in Fig. \ref{fig:mcleap-flowchart}. Two experimental assignments were given in two subsequent years (2017 and 2018), yielding four datasets with 426, 437, 346 and 399 participants respectively. There were only two students that participated in the assessments from both years. The assignment was designed to introduce a topic within the scope of the course that had not been previously taught (in this case reflection and refraction, which normally are taught toward the end of an introductory Electricity and Magnetism course). The pre-test is given directly after the instructions and consent form, and is meant to test the student’s prior knowledge as well as familiarize them with the assignment format. The pre-test consisted of five questions in both years. After the pre-test, the students completed a short questionnaire, asking them what type of content they would prefer to see as well as how prepared they feel going into the next assessment. After the first questionnaire, the students are shown a randomized piece of content from one of the three categories (theory, concept, or example), asked to solve two to four quantitative problems regarding the material they were just introduced to, and then fill out a similar questionnaire after to see if their preference or preparedness have changed. This was repeated twice more to cover all three types of content, then a longer (five questions in 2017 or ten in 2018) post-test was given. Student learning outcome was defined by their performance in quantitative problem-solving tasks as used in previous studies \cite{Andersson_2005}. Upon completion of the entire assignment, a final survey was given to gather information on student preferences, including which learning methods (for example use of resources) they found helpful.

The web interface used for this study allowed for the random selection of test questions and random content type pages, and could also generate content or questions based on previous answers. It tracked all of the answers as well as their timing. Students were required to complete one part of McLEAP in five hours, within a five day period. McLEAP consisted of two parts, each part testing different course content. The online structure of part 1 and 2 were the same. The test questions (which are automatically graded and provide immediate feedback to the students) are standard assignment questions (see Fig. S\ref{suppfig:quiz}) and the content pages are typical lecture content pages based on the three different content types (theory, example, and concept), which are illustrated in Fig. \ref{fig:illustration-content-type}. The entire McLEAP platform was written in PHP (a website programming language) and is stored on a McGill server. McLEAP stores all the answers in an encrypted data file (DATA) that can only be accessed by the PI (Hilke). From DATA we use a decoding program which generates two different files: FILE A, which contains the student IDs and their assignment grade. FILE B, which extracts the data from the students who have answered yes to the consent question, and assigns a random number to the remaining students and includes all the data saved by McLEAP. This ensures that analysis on FILE B can be done while keeping the participants anonymous.

\subsection{Analysis} 
\label{sec:analysis}

To quantify the effect of sequencing on learning outcome, we report normalized gain or just gain $g$. Gain was introduced by Hake \cite{hake1998} to assess whether ``the classroom use of Interactive Engagement methods increases the effectiveness of introductory mechanics courses well beyond that attained by traditional methods''. As Hake surveyed standardized test results from a range of institutions, he designed gain as a course-averaged measure reflecting learning outcome independent of pre-test results. It is defined as
\begin{equation}
\label{eq:gain}
g = \frac{ \langle G_\text{post} \rangle - \langle G_\text{pre} \rangle}{1 - \langle G_\text{pre} \rangle}
\end{equation}
where $\langle G_\text{post} \rangle$ and $\langle G_\text{pre} \rangle$ are the class-averaged post-test and pre-test results. As $g$ in Eq. \ref{eq:gain} does not allow one to investigate students' individual gain, Hake later introduced $\bar g$ \cite{hake2002}, the average of students' individual gains 
\begin{equation}
\label{eq:gain-bar}
\bar g = \left \langle \frac{  G_\text{post} - G_\text{pre}}{1 - G_\text{pre} } \right \rangle.
\end{equation}
For large classrooms the difference between gain $g$ and averaged gain $\bar g$ is small. Systematic differences between gain and averaged gain indicate what student profile profited most from the course \cite{bao2006}. $g - \bar g > 0$ means students with low pre-test scores had a higher gain, for $\bar g - g \geq 0$, the opposite is the case.

Despite over 20 years of wide use in physics education research (PER), several issues with gain or averaged gain are routinely raised. The first issue states that the gain of an individual is not a symmetric measure: while bounded from above, it is not bounded from below. This may induce a bias towards low pre-test scores. To address this issue, Marx and Cummings \cite{marx2007} proposed the normalized change, defined as 
\begin{equation}
\label{eq:norm-change}
  c =
    \begin{cases}
      \frac{  G_\text{post} - G_\text{pre}}{1 - G_\text{pre} } & \text{if }  G_\text{post} > G_\text{pre} \\
      \frac{ G_\text{post} - G_\text{pre}}{G_\text{pre} } & \text{if } G_\text{post} < G_\text{pre} \\
      0 & \text{if } 0 < G_\text{post} = G_\text{pre} < 1\\
      \text{drop} & \text{else} .
    \end{cases}       
\end{equation}
Normalized change may better take into account negative gain, but one may wonder what insights are recovered by correctly handling students with a high pre-score and a low post-score. Throughout the course, did these students unlearn the ability to answer questions correctly or did they coincidentally ace the pre-test \cite{coletta2020}? The solution proposed by Coletta and Steinert is to set a student's minimal gain to zero \cite{coletta2020}. Moreover, students with pre-test score $100\%$ will be discarded. A student's individual gain $g_\text{ind}$ under these constraints is computed as
\begin{equation}
\label{eq:gain-bar-1}
    g_\text{ind} =
    \begin{cases}
        \frac{  G_\text{post} - G_\text{pre}}{1 - G_\text{pre} } & \text{if }  G_\text{post} > G_\text{pre} \\
        0 & \text{if } G_\text{post} \leq G_\text{pre} < 1 \\
        \text{drop} & \text{else,} \\
    \end{cases}       
\end{equation}
and $\bar g$ follows by averaging over the student population. In this study, our ``course'' consists of a sequence of three content types and a total of five tests, which takes less than five hours to complete in total. It is not unreasonable to assume that some students will get confused and perform less well because of the sequence of content types. We thus take the stance that there is value in considering negative gain in this circumstance, which leads us to the second set of constraints for computing a student's individual gain
\begin{equation}
\label{eq:gain-bar-2}
    g_\text{ind} =
    \begin{cases}
        \frac{  G_\text{post} - G_\text{pre}}{1 - G_\text{pre} } & \text{if }  G_\text{pre} < 1 \\
        \text{drop} & \text{else,} \\
    \end{cases}       
\end{equation}
where again $\bar g$ follows by averaging over the student population. We report gain of averages $g$ (Eq. \ref{eq:gain}), normalized change $c$ (Eq. \ref{eq:norm-change}), and average of gains $\bar g$ computed via both methods (Eq. \ref{eq:gain-bar-1} and \ref{eq:gain-bar-2}), and remark on the differences. 

Finally, Nissen et al. \cite{nissen2018} argued that normalized gain is actually pre-score biased, although according to Coletta et al. \cite{coletta2020}, this is a misinterpretation of the findings of \cite{coletta2005}. They conclude that failure to consider other important predictors of gain like scientific reasoning ability (measured via Lawson's Classroom Test of Scientific Reasoning Ability \cite{lawson2000}) is an example of omitted variable bias and may lead to falsely concluding that gain is pre-score biased. We analyze the correspondence between gain and pre-test in our data, and note the importance of designing a good pre-test.

\section{Results} 
\label{sec:results-new}
The results are presented in two parts. First, we summarize the main result, analyze the effect of sequencing on four different measures of gain in detail, and study the variability between assignments and pre-score bias. Second, we discuss results from the student questionnaires during the assignment and from the independent survey on content type preference.

\subsection{Effect of Sequencing on Gain}
The first null hypothesis that we test is that \textit{no sequence results in a greater quantitative problem-solving ability than any other}. We find that the sequence CTE results in a significantly higher average of gains $\bar g$ than TCE ($p=0.03$), providing evidence against the null hypothesis (Fig. \ref{fig:gains-1}, left panel). Because this result was not supported ($p=0.08$) by gains of averages (Fig. \ref{fig:gains-1}, right panel), we decide to loosen our null hypothesis to \textit{no content type first  results in a greater quantitative problem-solving ability than any other}. The main finding of this paper, providing evidence against this null hypothesis ($p = 0.03$), is that Concept first leads to a significantly higher learning outcome than Theory first, a finding that is supported by both measures of gain (Fig. \ref{fig:gains-2}, left panels). The full explanation follows.

\subsubsection{No sequence provides a significantly better learning outcome than others}
\begin{figure}
    \centering
    \includegraphics[width=\columnwidth]{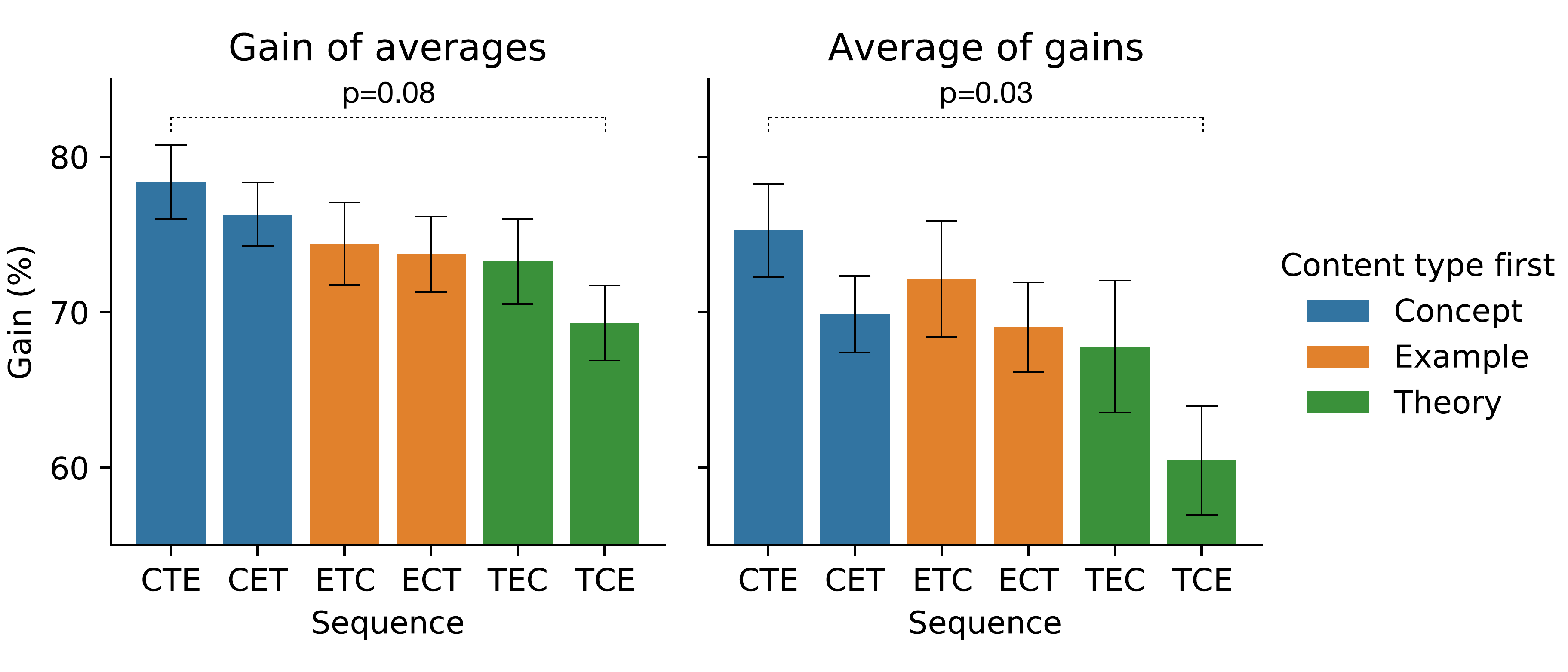}
    \caption{Gains per sequence averaged over four assignments. Dashed line indicates the hypothesis that CTE results in higher gain than TCE.}
    \label{fig:gains-1}
\end{figure}

Our null hypothesis states that \textit{no sequence results in a greater quantitative problem-solving ability than any other}. We measure quantitative problem-solving ability through various measures. First, we compute the normalized gain $g$ (Eq. \ref{eq:gain}) and the average of gains $\bar g$ (Eq. \ref{eq:gain-bar-2}) for each of the six sequences, visualized in Fig. \ref{fig:gains-1}. The error bars are standard errors of the mean. With Tukey’s HSD test, the p-value before reaching a significant result is adjusted by the number of hypotheses tested (15). We find that CTE gives a significant increase in average of gains TCE ($p=0.03$), but find a below significant result for gain of averages ($p=0.08$). We thus find no conclusive evidence against our null hypothesis.

\subsubsection{Concept first gives a significantly better learning outcome than theory first}
\begin{figure}
    \centering
    \includegraphics[width=\columnwidth]{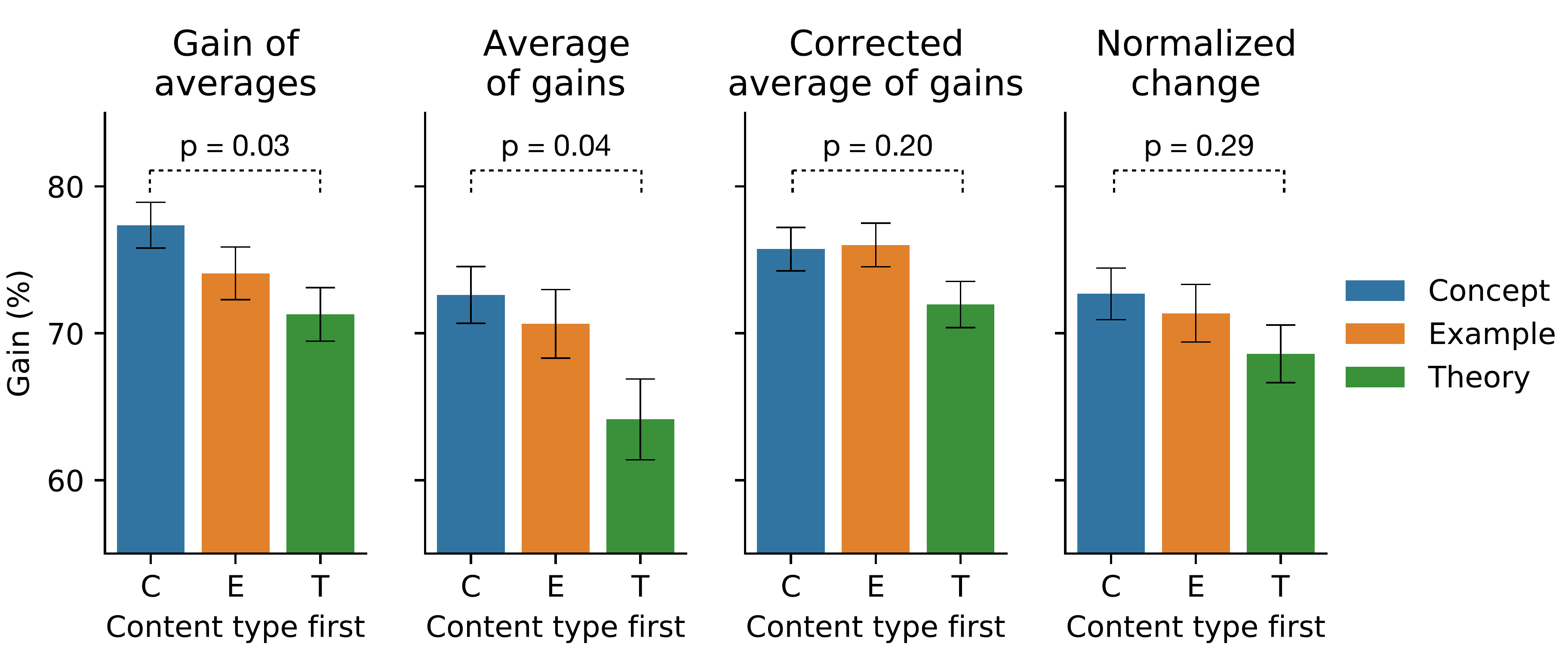}
    \caption{Gains per content type first averaged over four assignments. Error bars are standard errors. Dashed line indicates the hypothesis that concept first results in higher gain than theory first.}
    \label{fig:gains-2}
\end{figure}

Next, we loosen our null hypothesis to \textit{no content type first  results in a greater quantitative problem-solving ability than any other}. We chose to test content type first because students are most likely to be influenced at the start of the learning process due to declining attention with time \cite{bunce2010long}. The first content type thus carries most importance. Moreover, once a content type has been seen, it will not be repeated. In case the student's problem-solving ability is neither enhanced nor diminished following exposure to a content type, the student will also not be able to reap the benefits from a perhaps more advanced content type in a later stage of the learning process. Overall, this could provide a net negative effect compared to students who see this content type later, and gain from it. We again report normalized gain $g$ and average of gains $\bar g$ (Fig. \ref{fig:gains-2}) to test learning outcome based on content type first. Error bars are standard errors of the mean, and adjusted p values are computed with Tukey’s HSD with three hypotheses tested. Here we find a significant effect between Concept first and Theory first regardless of measure used ($p=0.033$ for normalized gain, $p=0.035$ for average of gains).

\subsubsection{Correcting for students with negative gain changes conclusion}
Before drawing conclusions based on significance in gain of averages and averages of gains, we wonder if the results are biased by the computation of these measures. As discussed in the section \ref{sec:analysis}, gain is not bounded from below when computed with Eq. \ref{eq:gain-bar-2}. Our findings for $\bar g$ may thus be biased towards contribution of strong negative gain. To validate this, we compute normalized change $c$ (Eq. \ref{eq:norm-change})
and corrected average of gains $\bar g$ where the minimum gain is set to zero (Eq. \ref{eq:gain-bar-1}) for subset students with the same content type first. We retrieve the same hierarchy for normalized change (Fig. \ref{fig:gains-2}) although the result is not significant (p = 0.24 for Concept first vs Theory first). For corrected $\bar g$, we find that students with Example first have a higher gain than students with Concept first (Fig. \ref{fig:gains-2}), neither of which is significantly higher than Theory first. How can taking into account negative gain differently so strongly influence our findings? 

To answer this question, we first count per content type the number of students with $G_\text{pre} > G_\text{post}$ excluding those with $G_\text{pre} = 1$ for whom we cannot compute gain. We then compute the mean gain and normalized change for this subset of students (Table \ref{tab:neg-gain}). Fewer students with Concept first (21) than Theory first (32) reported a negative gain, and students with negative gain in Concept first have on average a higher gain $(-76 \% \pm  13$) than those in Theory first ($-128 \% \pm 20$). Finally, the normalized change for these students in Concept first ($-45 \% \pm 6$) is higher than those with Theory first ($-55 \% \pm 6$), but the effect is less strong proportionally. Actually, the students with Example first have the largest mean negative normalized change ($-58 \% \pm 7$). This explains why we find no significance when we compute normalized change $c$ or corrected $\bar g$. But the question of which gain better represents the student’s learning outcome remains. Coletta et al. \cite{coletta2020} argued that over a semester-long course, it is unlikely that students unlearn the ability to solve problems. If they happen to score higher on the pre-test than on the post-test, this is most likely due to lucky guesses on the pre-test. The difference in pre-test and post-test then reflects no true (un)learning, therefore their argument is that the gain for these students should be zero. 

\begin{table}[!htb]
\caption{Students per content type first with $G_\text{pre} > G_\text{post}$ and their gain and normalized change}
\label{tab:neg-gain}
\begin{ruledtabular}
\begin{tabular}{cccc}
 & Number of students & $\bar g$ ($\%$) & $c$ ($\%$) \\ \hline
Concept first & 21 & $-76 \pm 13$ & $-45 \pm 6$ \\
Example first & 26 & $-103 \pm 16$ & $-58 \pm 7$ \\
Theory first  & 32 & $-128 \pm 20$ & $-55 \pm 6$ \\
\end{tabular}
\end{ruledtabular}
\end{table}

In our case study, the ``course'' lasted five hours or less and consisted of three learning moments through exposure to different content types, as well as through taking intermediate quizzes, for which results but not solutions are shown. It is not inconceivable that during this time-span, a subset of students reduced their problem-solving ability due to the sequence of content types. This is an argument in favor of measuring negative gain. Normalized change reflects this, but treats students with negative gain similarly. This reduces the significance of the result.

That 50\% more students with Theory first than with Concept first had a negative gain strengthens the statement that Concept first provides a better learning outcome. Exploring why this result is significant for $\bar g$ computed with Eq. \ref{eq:gain-bar-2} but not with Eq. \ref{eq:gain-bar-1}, we found that part of the significance is driven by students with a negative gain. These students were left off worse than they were before starting the course. It is unlikely for such a thing to happen during a one semester course, but for the shorter duration of our experiment, it is reasonable to assume this may indeed have occurred. These considerations show how various ways of computing gain provide insight into the learning process of subsets of students.

\subsubsection{Variability between assignments}
\label{sec:year}
We observe significant variability in pre-test results and gain between assignments and years. In this section, we will list factors contributing to this variability.

Pre-scores and post-scores for all four assignments are given in Table \ref{tab:test-per-test}. In 2018, pre-scores are $10\%$ higher than in 2017. The post-scores are also elevated, albeit less strongly. It caught our eye that in 2018, there were 69 students with $G_\text{pre}=1$, up from 8 students in 2017. We attribute this to students looking up questions and answers online using sources such as Chegg, demonstrating the limited durability of test questions for an online platform without external supervision on the use of resources. Correcting for students whom we cannot include in our calculation for $\bar g$ anyway, the results per assignment are given in Table \ref{tab:test-per-test-corrected}. This explains approximately $5\%$ of pre-score difference and does not affect the post-score difference. 

\begin{table}[!htb]
\caption{Mean grades and gain per assignment for all students}
\label{tab:test-per-test}
\begin{ruledtabular}
\begin{tabular}{ccccc}
assignment & pre-score ($\%$) & post-score ($\%$) & g ($\%$) & $\bar g$ ($\%$) \\ \hline
2017-1 & $39 \pm 1$ & $82 \pm 1$ & $70 \pm 2$ & $69 \pm 3$ \\
2017-2 & $31 \pm 1$ & $85 \pm 1$ & $78 \pm 2$ & $72 \pm 3$ \\
2018-1 & $50 \pm 2$ & $86 \pm 1$ & $72 \pm 2$ & $65 \pm 3$ \\
2018-2 & $41 \pm 2$ & $86 \pm 1$ & $76 \pm 1$ & $69 \pm 2$ \\
\end{tabular}
\end{ruledtabular}
\end{table}

\begin{table}[!htb]
\caption{Mean grades and gain per assignment for students with $G_\text{pre} < 100$}
\label{tab:test-per-test-corrected}
\begin{ruledtabular}
\begin{tabular}{ccccc}
assignment & pre-score ($\%$) & post-score ($\%$) & g ($\%$) & $\bar g$ ($\%$) \\ \hline
2017-1 & $39 \pm 1$ & $82 \pm 1$ & $70 \pm 2$ & $69 \pm 3$ \\
2017-2 & $29 \pm 1$ & $84 \pm 1$ & $78 \pm 2$ & $72 \pm 3$ \\
2018-1 & $44 \pm 2$ & $85 \pm 1$ & $71 \pm 2$ & $65 \pm 3$ \\
2018-2 & $35 \pm 2$ & $85 \pm 1$ & $74 \pm 1$ & $69 \pm 2$ \\
\end{tabular}
\end{ruledtabular}
\end{table}

Another difference is that the post-test contained ten questions in 2018, twice as many as in 2017. This did not affect the average post-scores, it only provided more resolution on the student's individual performance, and decreased the standard error from $1.4\%$ and $1.5\%$ (2017) to $0.9\%$ and $0.8\%$ (2018). 

A final factor that could have contributed to the enhanced 2018 pre-scores is the timing of the assignments. The 2018 assignments were given in week 13 and 14 of the course, while the 2017 assignments occurred both in week 12. The prolonged study period could have enhanced the student's prior knowledge. We do not know how large this factor is, as we did not do a pre-pre-test. Despite that the cohorts in 2017 and 2018 consisted of entirely different students (save two), the students had similar backgrounds (first-year engineering and science students mostly from Canada and the US). Moreover, the course structure was identical, as was the instructor. Yet, we still end up with unidentified variation. This demonstrates just how hard it is to properly execute a randomized control trial in PER \cite{lortie2019}. McLEAP may aid in creating randomized subsets of students, but the design of the study needs to be such that other sources of variation are accounted for as much as possible too.

In Table \ref{tab:test-per-test} and \ref{tab:test-per-test-corrected} we have included gain for the full and corrected population and for the corrected. $\bar g$ is not affected by the correction, because students with $G_\text{pre} = 1$ were never included in computing $\bar g$. Different from the pre-scores and post-scores, gain of averages $g$ is generally preserved, independent of the inclusion of potentially cheating students. In terms of Hake's classification scheme \cite{hake1998}, the gain is large for all assignments ($ 70\% - 100 \%$), because of the lack of previous experience students had with solving such problems. The normalized gain seem to be negatively correlated with pre-scores, an observation that we will explore in the next section. 

\subsubsection{Pre-score bias}
\begin{figure*}[t]
\includegraphics[width= \textwidth]{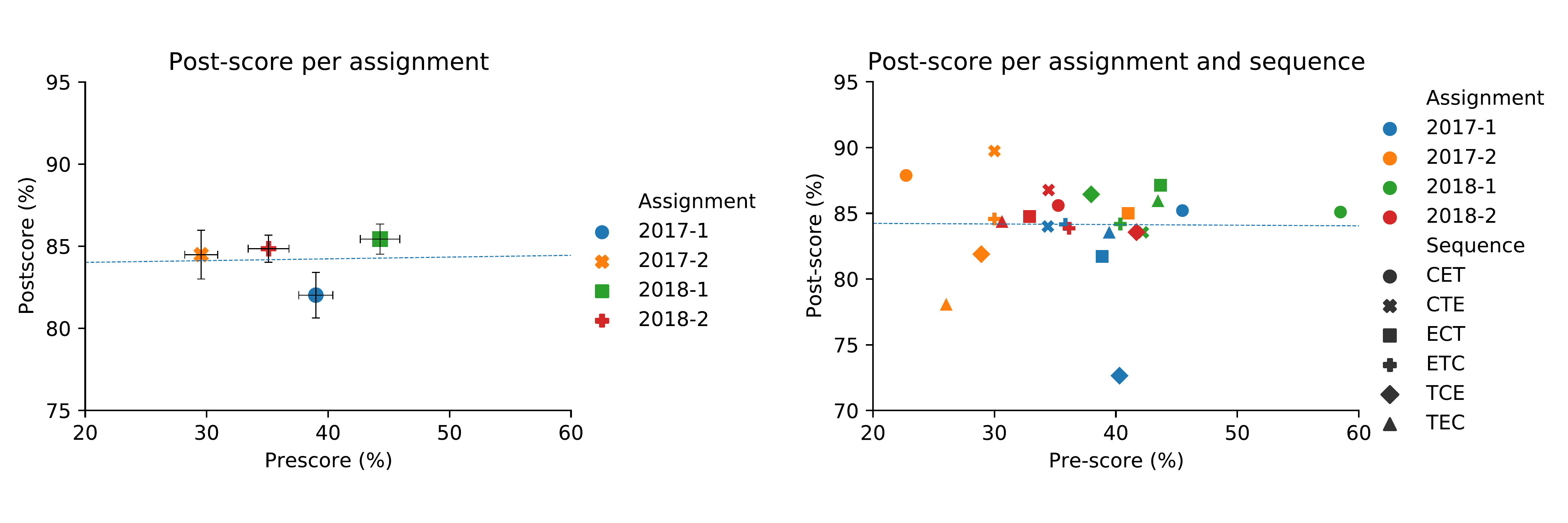}
\caption{Post-scores as a function of pre-scores per assignment (left) and per assignment and sequence (right). Error bars are standard error of the mean. Dashed blue line shows the fit with coefficient $0.0 \pm 0.2$, meaning that post-scores are independent of pre-scores.}
\label{fig:post-pre}
\end{figure*}

\begin{figure*}[!tb]
\includegraphics[width=\textwidth]{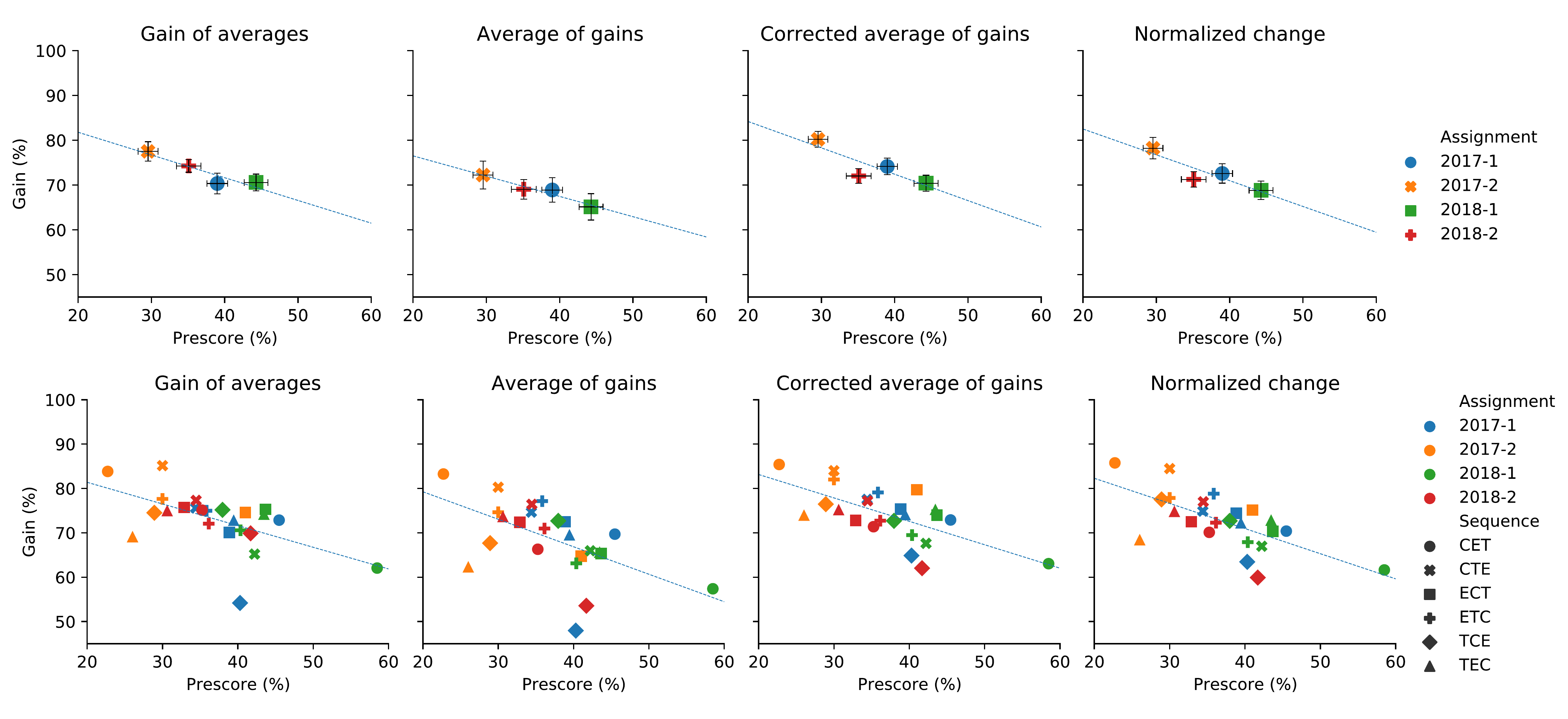}
\caption{Gain of averages, average of gains, corrected average of gains and normalized change as a function of pre-score per assignment (top panels) and per assignment and sequence (bottom panels). Error bars are standard error of the mean. Dashed blue line shows the fit with coefficients $-0.5 \pm 0.1, -0.5 \pm 0.1, -0.6 \pm 0.3, -0.6 \pm 0.2$, meaning that normalized gain strongly depends on pre-scores.}
\label{fig:gain-pre}
\end{figure*}

Normalized gain was introduced as a measure of student learning independent of pre-score \cite{hake1998}. In this section, we analyze this dependence in our data. First, we note that post-scores are pre-score independent whether the data is averaged per assignment (left panel of Fig. \ref{fig:post-pre}) or per assignment and sequence (right panel of Fig. \ref{fig:post-pre}). On the left panel, the error bars are standard errors, on the right panel, every point can be considered a course, and to avoid confusing, we do not plot the many error bars, which are a bit larger than in panel A. With post-scores independent of pre-scores (coefficient $0.0 \pm 0.2, \, p = 0.96$, we are actually introducing a pre-score bias in normalized gain. This can be seen in Fig. \ref{fig:gain-pre}. For the four measures of gain, we plot their pre-scores per assignment (top), and per assignment and sequence (bottom). Coefficients are given by $-0.5 \pm 0.1, -0.5 \pm 0.1, -0.6 \pm 0.3, -0.6 \pm 0.2$, meaning that we have introduced a robust dependency on gain independent of the measure used to compute gain. In Figs. \ref{fig:post-pre} and \ref{fig:gain-pre}, we excluded students with $G_\text{pre}=1$.

Post-scores should depend on pre-scores; it is a fact of life that students with more prior knowledge do better in post-tests. That the pre-test did not adequately capture the student's prior knowledge may be due to the test design or due to external conditions, i.e. because ungraded students may not have taken the test seriously or a majority of students looked up questions online, an effect still visible with students with pre-score $100\%$ excluded. It could also be that we should have tested for prior skills through mathematical aptitude, instead of testing prior domain knowledge. In 2018, more than a month before McLEAP, students completed several mathematical oriented assignments (math tests), consisting of seven questions on topics like vector algebra, differentiation and integration. Multivariate regression on the data excluding students with the maximal pre-score shows that the score on the math test (coefficient $0.13\pm0.04$) is a better predictor of the post-score than pre-score (coefficient $0.03\pm0.02$). These coefficients remain preserved when predicting post-score through a univariate regression with pre-score alone ($0.04\pm0.02$) or math-score alone ($0.13\pm0.04$), showing that the pre-test and the math test measure different skills. Finally, the difference in regression coefficients would be even larger in 2017, assuming that the math-score coefficient is preserved across years, because the pre-score has an even smaller dependence on post-score here (Fig. \ref{fig:post-pre} left panel). 

Previously, Coletta et al. \cite{coletta2020} refuted criticism by Nissen et al. \cite{nissen2018} on pre-score bias in normalized gain by arguing that pre-score does not introduce a bias if there are unmeasured variables that influence the pre-score. In their case, scientific reasoning ability measured by Lawson scores \cite{lawson2000} was the confounding variable that influenced both pre-score and normalized gain. In our case, math score is not a confounding variable of pre-score, but is a stronger predictor of post-score. Had we used these math-test as our pre-score, we might have found a normalized gain that was less dependent on pre-score, while the post-test had a stronger dependence on pre-test. To complete the analysis, we compute post-scores for each of the sequences and content types first (Fig. S\ref{suppfig:post-test}), resulting in analogous findings when computing gain. Post-scores of students with the sequence CTE are on average $5\% \pm 3$ higher than students with TCE with multiple-hypothesis adjusted p-value 0.10. Post-scores of students with Concept first are on average $4\% \pm 2$ higher than students with Theory first, p-value 0.01. For completeness, we present the post-scores per content type first and per assignment in Fig. S\ref{suppfig:post-per-dataset}. The conclusions that Concept first gives better results than Theory first holds in three of four assignments separately. In 2018-1, the availability of questions online may have caused this statistical outlier. 

We conclude this section with the observation that the pre-test is an important part of any educational piece of research, but despite not being able to capture prior knowledge well, our results, coming from a randomized experiment with over 1500 datapoints, still hold. In this work, we have demonstrated that over a short learning period of less than five hours, showing students concept-related material first results in a greater quantitative problem-solving ability compared to showing students theory-related material first. 

\subsection{Sequence preference}
\label{sec:preference}
Sequencing matters, how is it viewed by students? After discussing student's content type preference during the experiment, we compare it more broadly to academics with varying levels of expertise (students to faculty) using an independent survey on content type preferences. 

\begin{figure*}
\includegraphics[width=\textwidth]{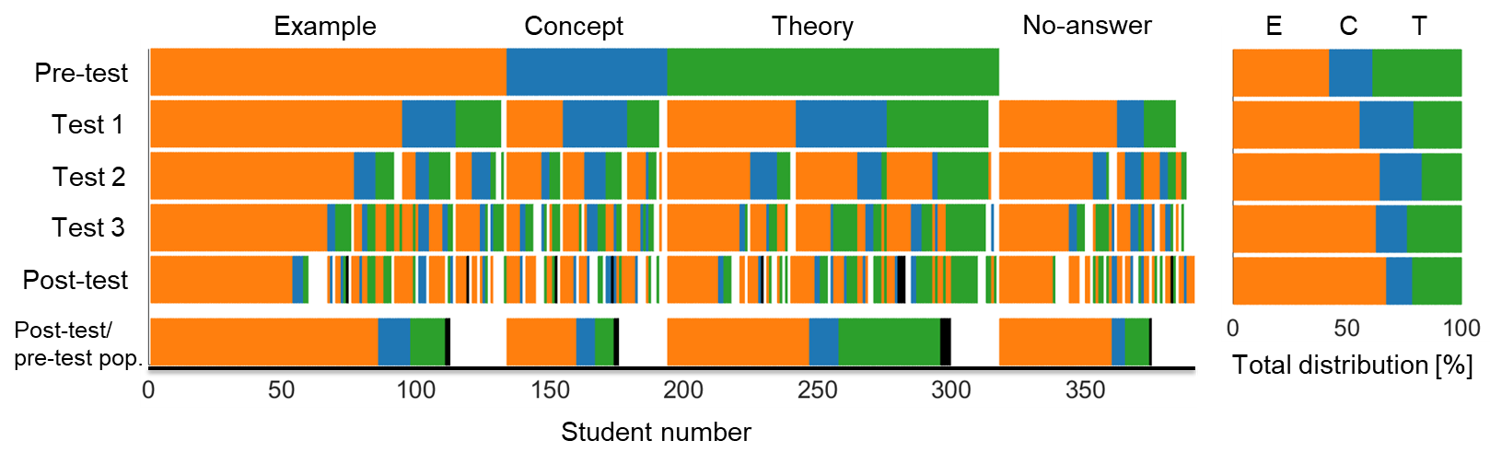}
\caption{Details of the preference change by student. The student preferences Example (E), Concept (C) and Theory (T) are represented by the colors orange, blue and green, respectively. White is used to represent a lack of answer, while black is used for more than one preference. The top row is the initial pre-test preference, while the lower rows show subsequent tests. The lowest row represents the post-test preference distribution based on the initial pre-test student preference population. The right column shows the relative preference evolution over the course of the different tests. Overall, the preference evolves towards example for all sub-groups, regardless of initial preference.}
\label{fig:preference}
\end{figure*}

Fig. \ref{fig:preference} shows results of the questionnaire for content type preferences over time. Every thin vertical slice represents a student, the colors indicate which content type the student prefers to see next. We find that the percentage of students who prefer Example rises from 42\% before Test 1 to 66\% after the post-test, the preference of Theory falls from 39\% before Test 1 to 22\% after post-test, and the preference of Concept falls from 19\% before Test 1 to 12\% after the post-test. Students change content type preference over time, \textit{i.e.} their preference of content order is developmental; they are improving and building upon their previous conceptions throughout the assignment. As the post-test approaches, students prefer the example content type. This may point towards a shift away from understanding the phenomenon towards learning how to solve problems. This preference is justified, given that  students who were assigned the example content type first scored $2\%$ higher on Test 1 than students who were assigned the concept first, and $3\%$ higher than students with theory first. These results should be taken with caution as intermediate tests consisted of no more than three questions. This phenomenon is called ``surface learning'' \cite{dinsmore2012}, as by examples only, students may find a learning shortcut that helps on shorter timescales, but will not be able to acquire the deep understanding that is required achieve the best of their abilities on long timescales, as sequences with concept first offer. In future studies, it may be of interest to loosen the constraint that all content types have be to seen once. We hypothesize that a sequence of three non-unique example content types will still not give better problem-solving results than the concept first sequences.

\begin{figure}[h]
\includegraphics[width=\columnwidth]{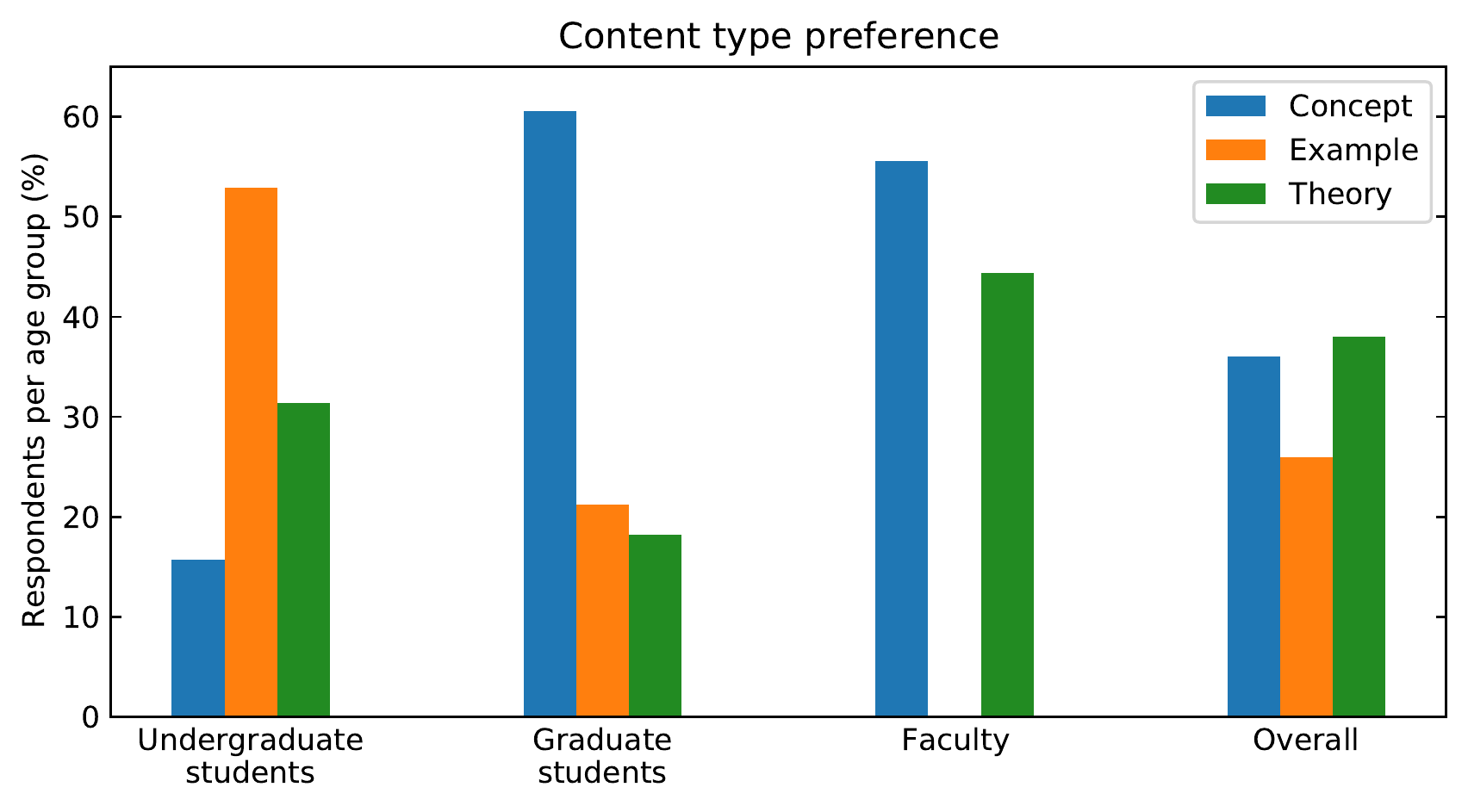}
\caption{Survey results for preferred content type used for self learning for undergraduate student, graduate students, faculty members and overall survey respondents}
\label{fig:survey-preference}
\end{figure}

It should be noted that for these questionnaires amidst the McLEAP assignment, students are obviously focused on acquiring problem-solving skills, which might skew their content type preference. To find out about the ``true'' or long-term preference we conducted an independent survey inquiring undergraduate students, graduate students, and faculty members about content type and sequence preference and when learning new material. Results for the content type preference are given in Fig. \ref{fig:survey-preference}. With experience, content type preference shifts from example (undergraduates) towards concept (graduates), and a mix of concept and theory (faculty), consistent with the view that teachers may have beliefs on teaching science that are more traditional than constructivist \cite{tsai2002}.

\begin{figure}[h]
\includegraphics[width=\columnwidth]{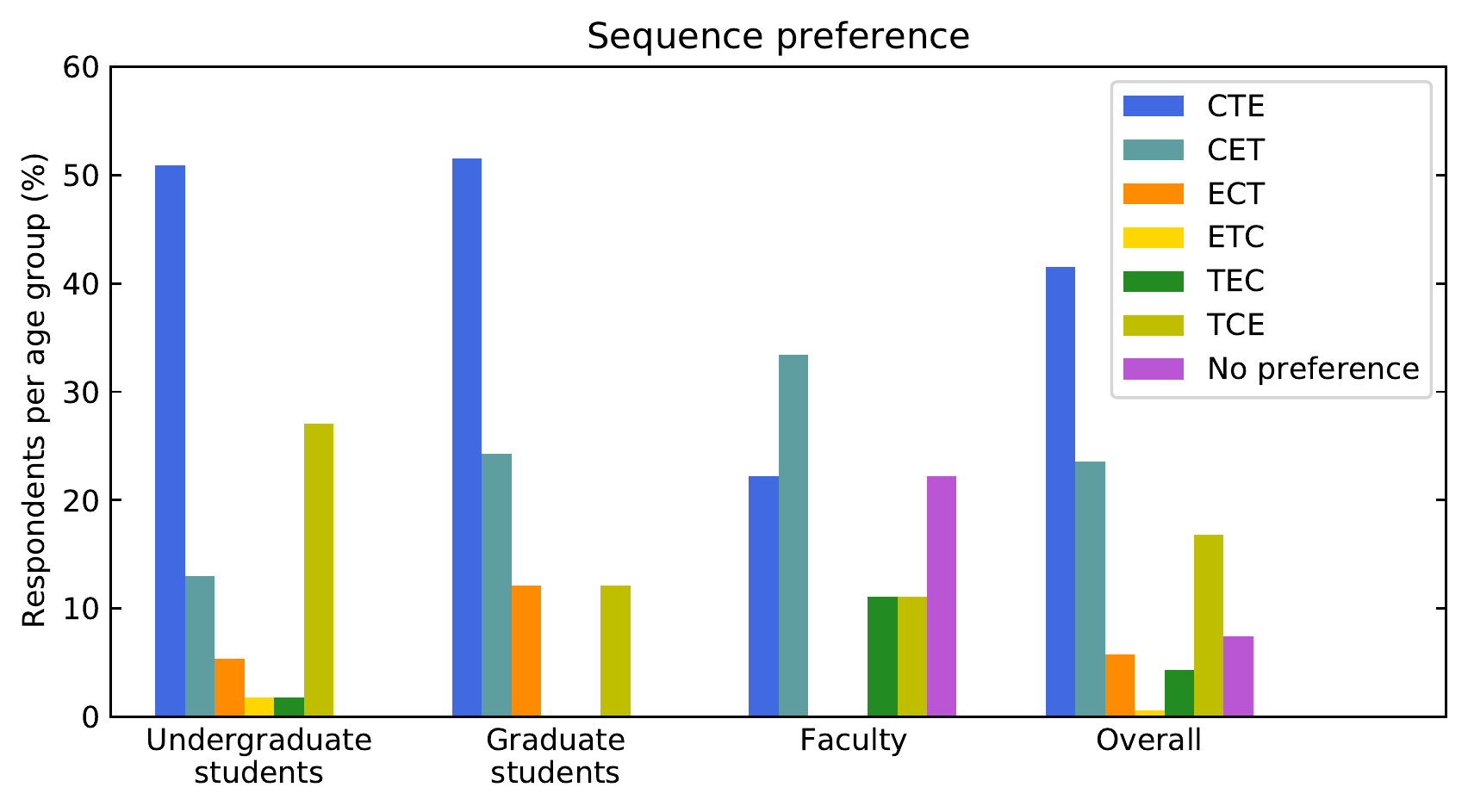}
\caption{Survey results for the question: ``Is there an order for presenting educational material that you think is best for understanding new concepts?". Percentage of respondents who preferred each permutation of content order, grouped by the class of the participants.}
\label{fig:survey-order}
\end{figure}

To see how the content preference relates to the sequence preference, we also asked the respondents about their sequence preference (Fig. \ref{fig:survey-order}). Interestingly, the sequence preference corresponds well to the sequences with which students performed best during the McLEAP assignments. This is in contrast to Kohl and Finkelstein \cite{kohl}, where they found that students' sequencing preference did not align with the questions they performed best at. The main difference between our studies was that Kohl and Finkelstein worked with pictorial and graphical content types, while our content types were provided mostly in written style. Using pictorial content, forces students to use representational skills at many levels, which is different when they only use written content.

We conclude this section with the observation that the sequencing preference was asked in terms of learning new material over an unspecified timescale, while the optimal sequence was discovered using a ``course'' that lasted less than five hours. Yet, one could consider a semester-long course consisting of many short term courses about unique topics, and that for each of the individual topics, the best way of structuring the learning procedure is through a concept-first approach. Our results therefore align more with the constructivist theory of learning in that learning is not only about exposure to different content types, but also about the existence of general rules on what content type should be seen first, especially in the case of avoiding the introduction of incorrect understandings or interpretations. We are exploring additional studies probing the effects of misconceptions using the McLEAP tool.

\subsection{Preparedness}

In the previous section we found an alignment between the content preference and the learning gain. Here we extend this analysis to the preparedness or self-efficacy. For the 2018 part II data, we included questions not only on the preference (section \ref{sec:preference}), but we also on preparedness, measured with survey questions throughout the experiment.

We find that students presented with concept first experienced the most positive percentage change in preparedness overall from Test 1 to the Posttest, compared to students who were presented with example and theory first. We use the student’ self-reported preparedness as a marker for student confidence and self-efficacy \cite{dowd2015making}. Previous studies have found that various forms of educational and motivational scaffolding, as well as effective learning strategies, increase students' self-efficacy which has been linked to improved learning achievement  \cite{wadsworth2007online,yantraprakorn2013enhancing,prabawanto2018enhancement,valencia2018effect}

\begin{figure}[!htb]
\includegraphics[width=\columnwidth]{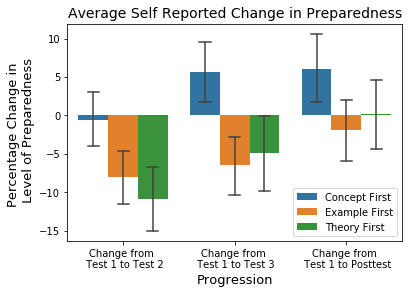}
\caption{Self reported preparedness from Test 1 to each level of the assignment. The confidence interval is 95\%. Students are asked to rate their preparedness from 1 (low) to 5 (high) at each stage of the assignment, and we extrapolate the percentage change in preparedness over time with this data.}
\label{fig:changeinpreparedness}
\end{figure}

As shown in Fig. \ref{fig:changeinpreparedness}, we find that the percentage change in preparedness of the students in the concept first groups is -0.5 $\pm $ 1.8 from Test 1 to Test 2, 5.6 $\pm 2.0$ from Test 1 to Test 3 and finally 6.1 $\pm 2.2$ overall from Test 1 to the Posttest. The students who receive concept first do not report an overall change in preparedness that is statistically significant within a 95\% confidence interval as it compares to the overall change in preparedness within the example and theory first groups. Concept first is however the only group that significantly increases in preparedness over time. The students’ preparedness beginning with concept first agrees with the previously reported results, where the order that yields the greatest learning outcome is also concept first. Moreover, the results of this analysis on preparedness suggest that students are aware of what type of content, and which content order, is allowing them to acquire enough knowledge to properly understand the material at each step in the assignment.

\section{Conclusions}
\label{sec:conclusion}

%Sequencing matters (concept first), Preferences (dynamic in learning process and dynamic in learner's, preparedness.
{\bf Main results:} The results of our two-year study with close to 1500 students completing the McLEAP assignment show that sequencing of content types has a significant impact on quantitative problem-solving measured by normalized gain. We find that there is an optimal sequencing of material presentation that leads to greater problem-solving performance. The optimal sequence of content presentation is {\it Concept first}, while a significantly worse outcome resulted when {\it Theory first} is presented. The advantage of {\it Concept first} vs. {\it Theory first} is statistically significant (p$<$0.05) over different measures, be it the gain of averages, the average of gains or the final post-test scores. The gain for {\it Example first} lies in-between {\it Concept first} and {\it Theory first}, but without statistical significance.

These results have {\bf important implications} for both the Learner (L) and the Teacher (T), and it is interesting to look at these two perspectives separately. For T, with a given collection of teaching content, it not only implies that the order in which material is presented matters, but it also prescribes the optimal sequencing of the learning material in this particular case, which is teaching topics in undergraduate Electromagnetism. While "most faculty in higher education initially adopt a teaching style that merges (1) the ways they prefer to learn and (2) approaches to teaching they saw as effective for their own learning in their higher education programs" as quoted from \cite{hawk2007using}, our experiment provides a direct answer to the optimal sequencing of content, which could and should be adapted to different teaching environments.

% An issue with normalized gain is that it is known to correlate with gender \cite{hake2002,willoughby2009} and race \cite{vandusen2020}. Not having collected personal data during our study, we cannot comment on this, but we stress that this is an important issue that requires further study.

It is tempting to {\bf explain} the effect of quantitative problem-solving ability with content sequencing by the learner's preference. Indeed, several authors have argued that learning style should match a learner's preference, the "meshing hypothesis" \cite{pashler2008learning,sankey2011impact}, though no clear evidence has emerged that learning preference correlates with learning success \cite{aragon2002influence}. Indeed, while we observe differences in the content sequence preference as well as changes in this preference, we do not find any significant difference in the learning outcomes. This is in contrast to the ``surface learning'' trap that is suggested by intermediate test results. Many educators have likely experienced the common request by students that they want more examples when learning new material. We find that the initial gain from seeing an example first is only short-lived. Indeed, when students see all three content types, the highest gain is achieved when {\it Concept first} is presented and not {\it Example first}. 
We also find a stark difference between the sequence preference of undergraduate students versus faculty, further illustrating that preference in sequencing is not static. While preference might not affect learning outcome it could nevertheless affect a student's overall satisfaction \cite{witowski2008relationship}. 

Student readiness or preparedness is sometimes associated with academic performance in the literature \cite{le2005motivational}. For our main result on the best learning gain, which favors Concept first, we also find that the change in preparedness is greatest with Concept first, suggestive of an association between preparedness and academic performance.

Can we go further in our understanding of the significant differences in the learning gain with different content sequencing? Our experiment draws on the constructivist framework within the theory of learning. While we have explored several aspects related to this framework, such as Bloom's taxonomy, constructivist and generative theories of learning, our results can be best visualised by thinking of three bricks. The largest is C, the middle sized brick is E and the smallest one is T. Building by starting with C is the most robust in analogy with gain, while starting with T is unstable. Building using only E bricks works initially and allows for quick results, but ends up being less robust overall. This constructivist visualization also draws on the concepts of scaffolding theories, where our choice of different content styles have different scaffolding qualities and add a time axis to the scaffolding. 

This is a {\bf case study} with students learning Electromagnetism using an online platform in their first year of university. While this is a relatively limited scope, one can easily imagine that similar studies can be performed in {\bf other STEM subjects} or even beyond. The key element is content structuring. While we chose a structuring along E, C and T, there can be other structures, like for example, verbal, mathematical, graphical, or pictorial \cite{kohl}. The structuring will depend on the topic, focus of study as well as level of study. While it is readily possible to translate our study to different levels (high school or even graduate studies), it is less clear how it impacts in-class teaching, where the nature of the delivery is harder to control. Our McLEAP online platform allows for a more uniform delivery, independent on the chosen structuring, which will allow for future studies at other levels and different fields.

\begin{acknowledgments}
We would like to thank Chris Roderick, Janette Barrington, Marcy Slapcoff, V\'{e}ronique Brul\'{e}, and Rebecca Brosseau for useful discussion, editing, and literature recommendations. We would like to thank Zezhou Liu for programming support on the online learning platform.
\end{acknowledgments}

\bibliographystyle{unsrt}
\bibliography{apssamp}

\end{document}

\captionsetup[figure]{name={FIG. S},labelsep=period}
% \section{\captionsetup[table]{name={TABLE. S},labelsep=period}}
\captionsetup{justification=raggedright,singlelinecheck=false}

%\title{Supplemental Material}

%\author{Benjamin J. Dringoli}
%\author{Ksenia Kolosova}
%\author{Thomas J. Rademaker}
%\author{Juliann Wray}
%\author{J\'{e}r\'{e}mie Choquette}
%\author{Michael Hilke}

% \title{Supplementary Material for Content Sequencing and its Impact on Student Learning in Electromagnetism: Theory and Experiment}

\title{Supplementary Material for The impact of content sequencing on quantitative problem solving: A case study in Electromagnetism using an online learning platform}

\author{Benjamin J. Dringoli}
\affiliation{Department of Physics, McGill University, 3600 rue University, Montr\'{e}al, Qu\'{e}bec H3A2T8, Canada}
\author{Ksenia Kolosova}
\affiliation{Department of Physics, McGill University, 3600 rue University, Montr\'{e}al, Qu\'{e}bec H3A2T8, Canada}
\author{Thomas J. Rademaker}
\affiliation{Department of Physics, McGill University, 3600 rue University, Montr\'{e}al, Qu\'{e}bec H3A2T8, Canada}
\author{Juliann Wray}
\affiliation{Department of Physics, McGill University, 3600 rue University, Montr\'{e}al, Qu\'{e}bec H3A2T8, Canada}
\author{Jeremie Choquette}
\affiliation{Department of Physics, McGill University, 3600 rue University, Montr\'{e}al, Qu\'{e}bec H3A2T8, Canada}
\affiliation{Dawson College, 3040 Sherbrooke St W, Montr\'{e}al, Qu\'{e}bec H3Z 1A4, Canada}
\author{Michael Hilke}
\affiliation{Department of Physics, McGill University, 3600 rue University, Montr\'{e}al, Qu\'{e}bec H3A2T8, Canada}

%\date{\today}% It is always \today, today, but any date may be explicitly specified


\maketitle

\section{Supplementary Figures}

\begin{figure}[!htb]
\includegraphics[width=0.5\columnwidth]{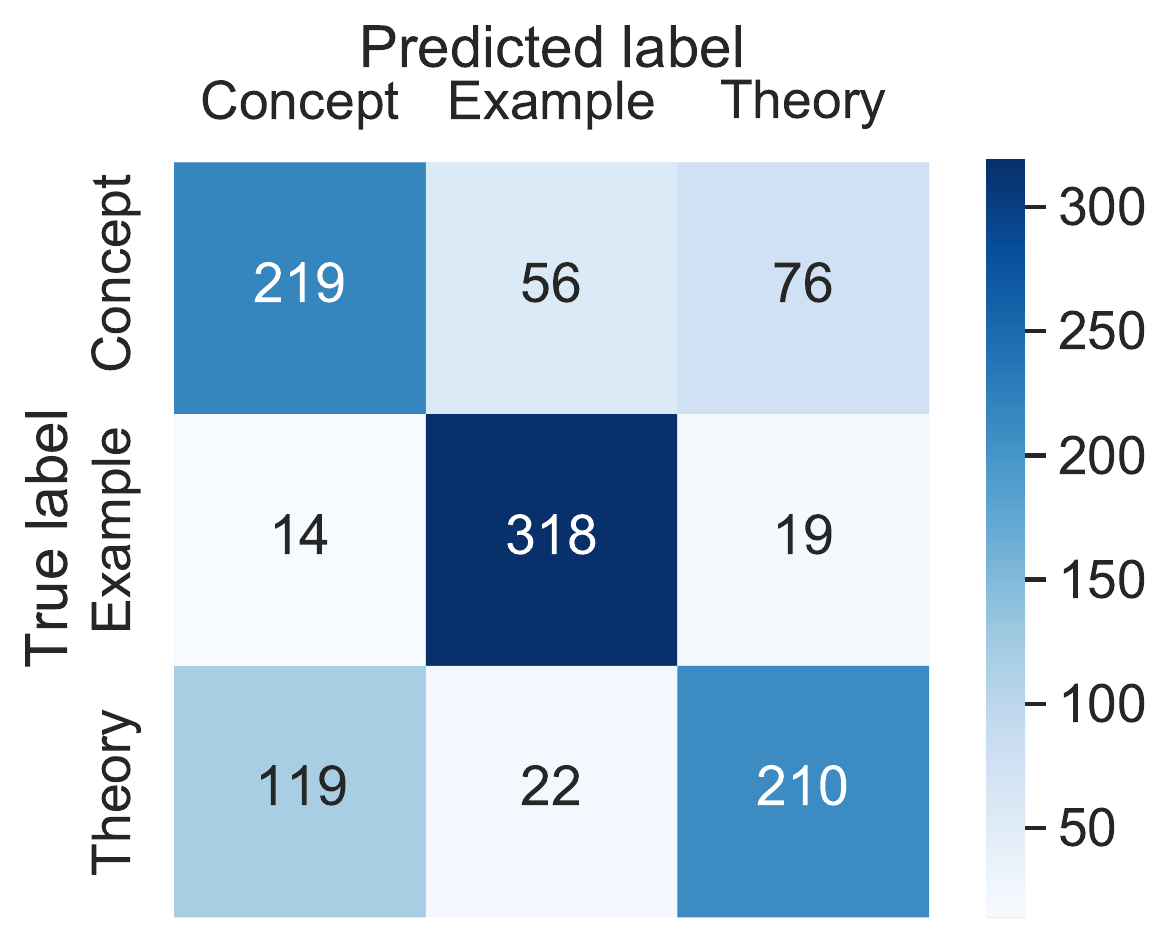}
\caption{Survey results for the question \textit{Choose the description you think best describes the educational content shown} for 9 pieces of content. Overall, a majority of survey respondents accurately label each content type with the same categorization that we use in this experiment.}
\label{suppfig:accuracy-labelling}
\end{figure}

\begin{figure}[!htb]
\includegraphics[width=\columnwidth]{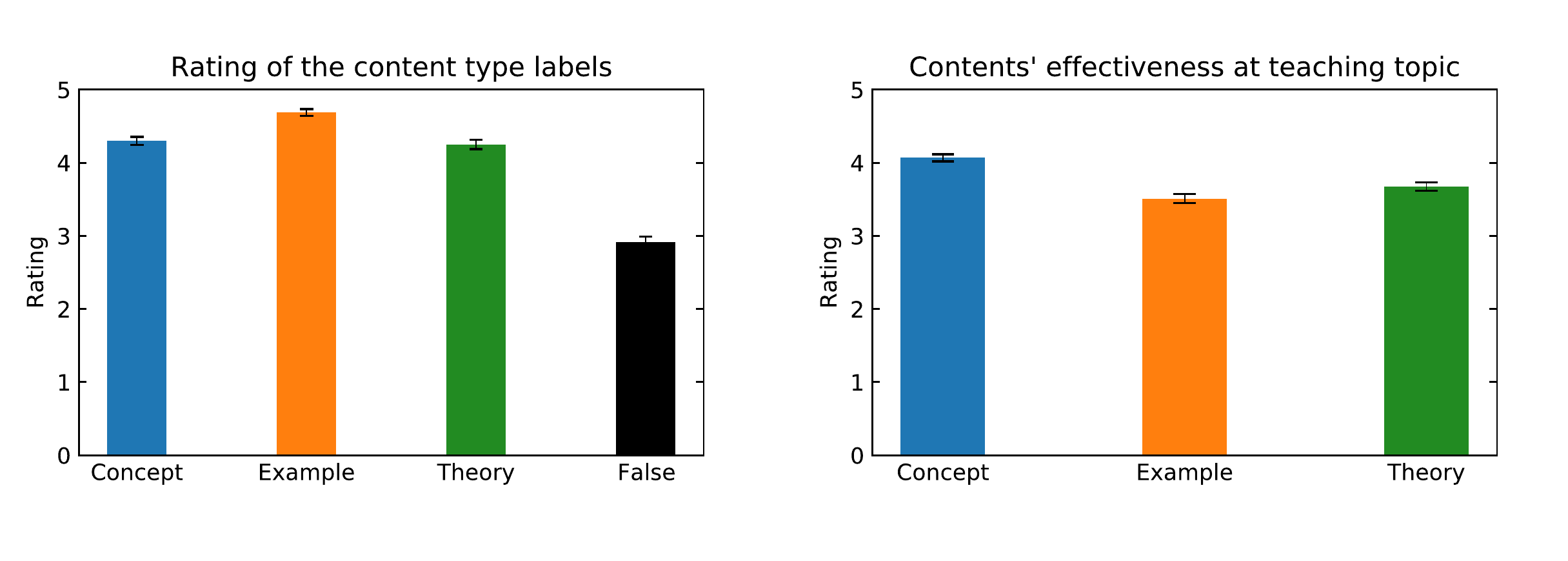}
\caption{Survey results for rating the accuracy of the content type label (left panel) and its effectiveness in teaching the topic (right panel) for 9 pieces of content. On the left, survey respondents answered the question \textit{How accurate does the label 'Content type' fit for the content above?} Content type was either Concept, Theory or Example, and accuracy was measured on a scale from 1 to 5, 1 being very inaccurate and 5 being very accurate. We find a rating higher than 4 out of 5 for content type that was correctly labeled (first three bars), and distinctly worse ratings for the falsely labelled content (right bar). On the right, survey respondents answered the question \textit{How would you rate the quality of the content above in explaining ‘a physical concept’?} Concept has the highest rating, but Example and Theory are close together, demonstrating that learners are able to learn the topic from each of the content types. Error bars represent standard error of the mean.}
\label{suppfig:rating-content}
\end{figure}

\begin{figure}[!htb]
\includegraphics[width=0.9\columnwidth]{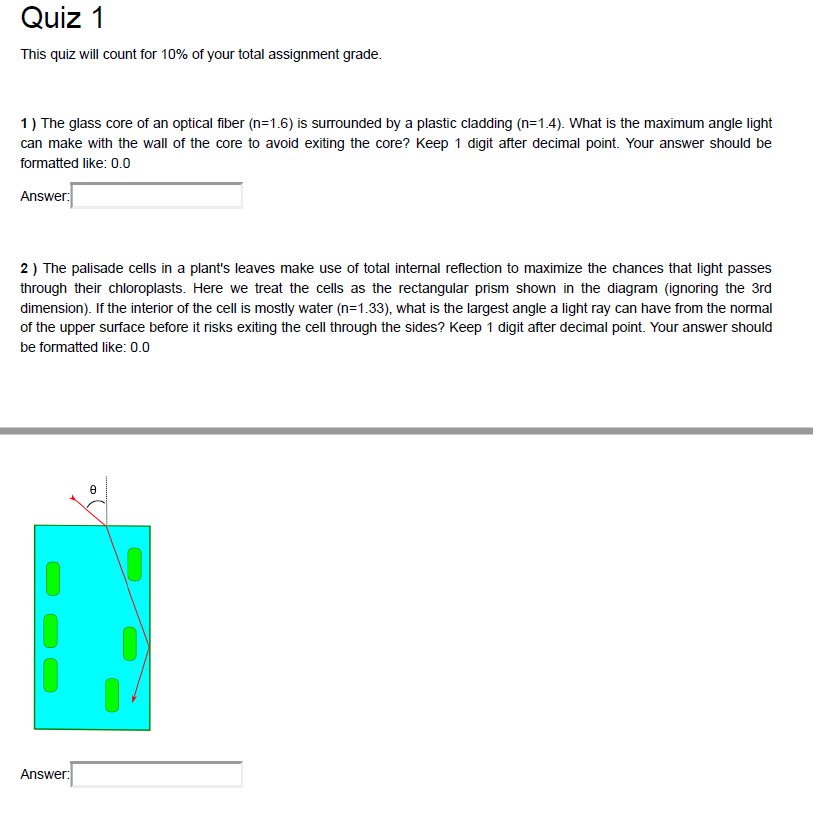}
\caption{Student view of McLEAP showing Test 1.}
\label{suppfig:quiz}
\end{figure}

\begin{widetext}
\begin{figure}[!htb]
\includegraphics[width=0.9\textwidth]{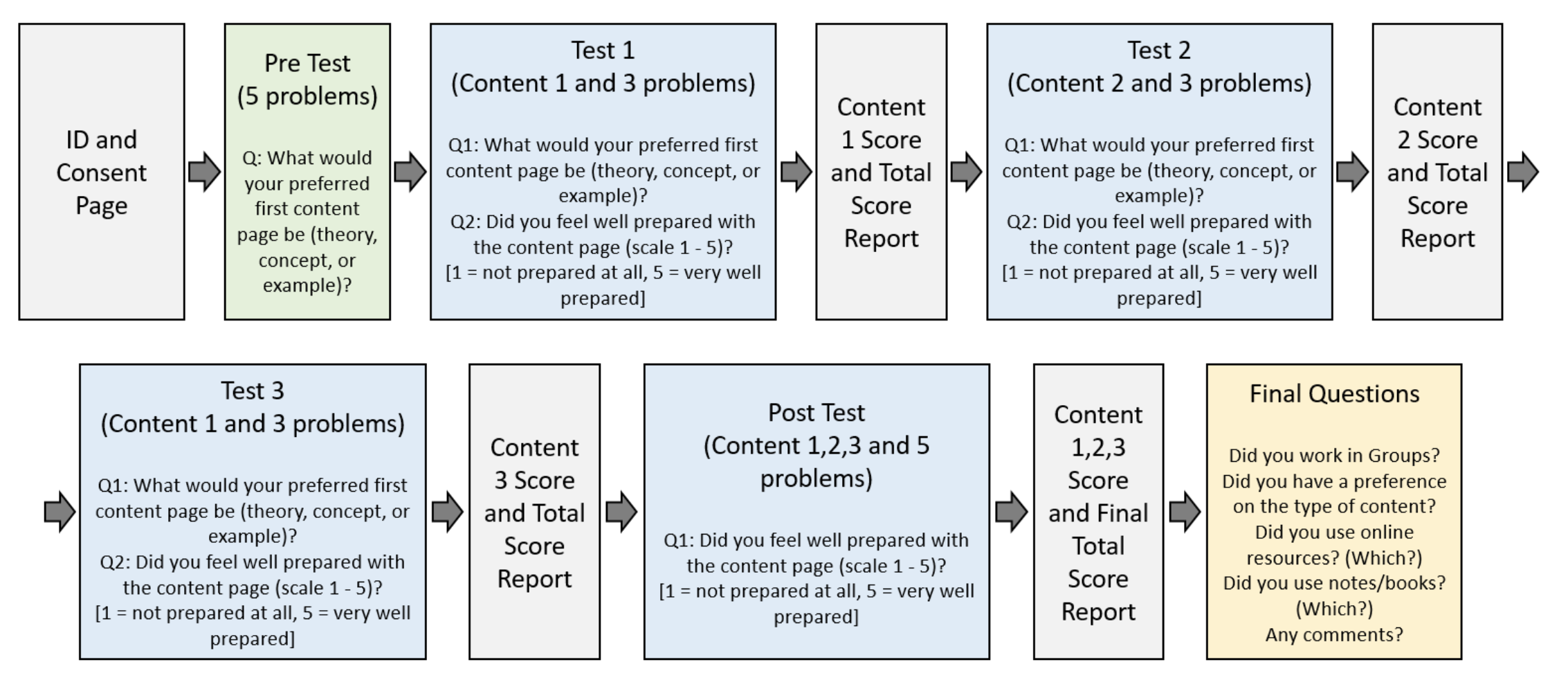}
\caption{Pages diagram of the McLEAP assignment interface; shows student progression through the assignment with the order of presentation for both problems and questions. Each arrow represents a new webpage seen by the students.}
\label{suppfig:mcleap-interface}
\end{figure}
\end{widetext}

\begin{figure}[!htb]
\includegraphics[width=0.8\columnwidth]{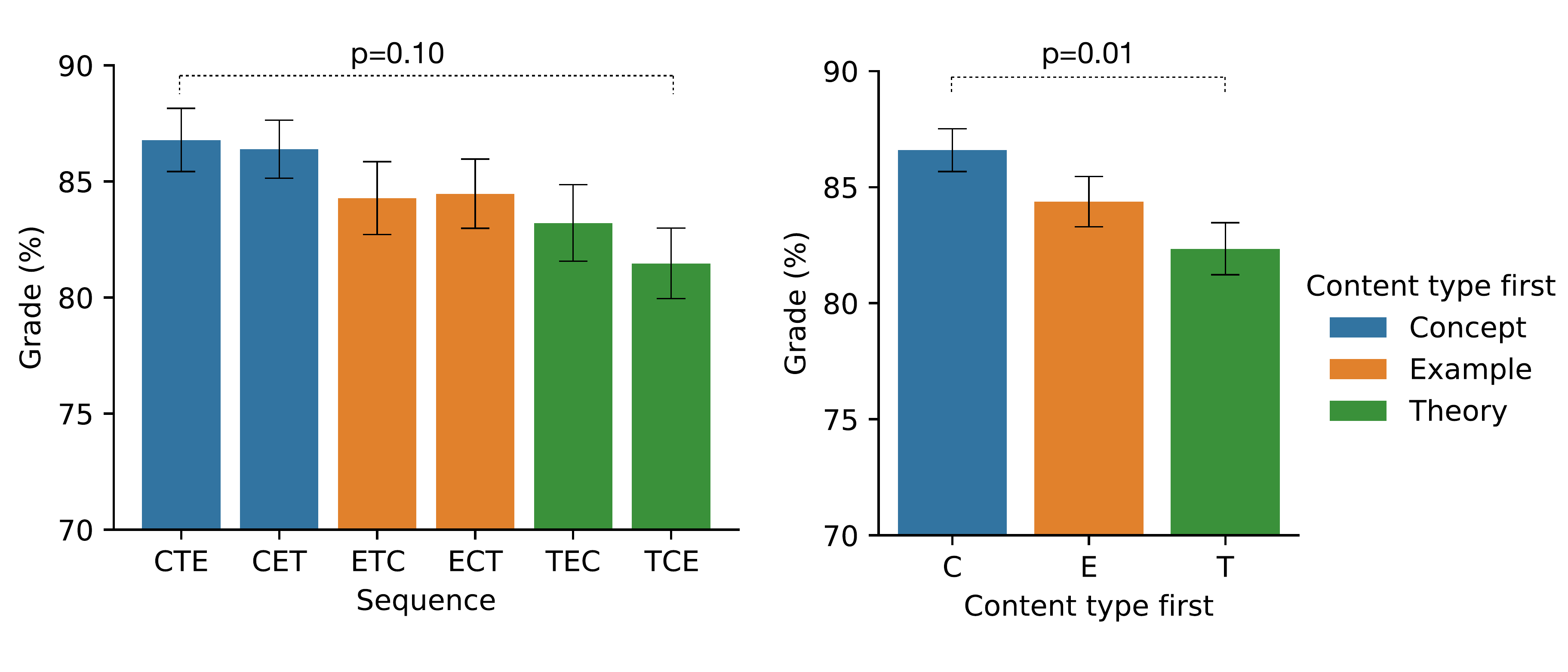}
\caption{Post-scores grouped per sequence (left) and content type first (right) averaged over four assignments. Dashed line indicates the hypothesis that CTE (concept first) results in higher gain than TCE (theory first).}
\label{suppfig:post-test}
\end{figure}

\begin{figure}[!htb]
\includegraphics[width=0.6\columnwidth]{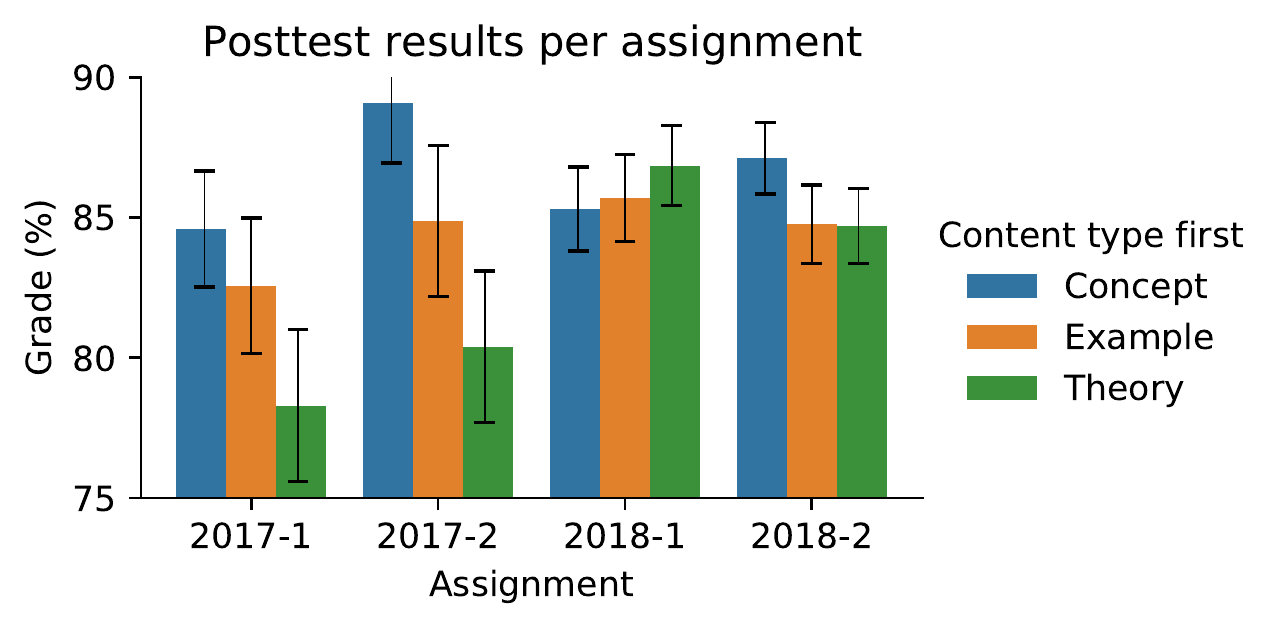}
\caption{Post-scores grouped per content type first and assignment (Data 2017-2018). Error bars show standard error from the mean. They are higher in Data 2017, because only 5 instead of 10 questions were asked during the posttest. Three out of the four data sets follow the same trend. Part 1 2018 is interpreted as a statistical outlier (no significant trend).}
\label{suppfig:post-per-dataset}
\end{figure}

This section contains Supplementary figures. Figs. S\ref{suppfig:accuracy-labelling} and S\ref{suppfig:rating-content} show more details on the survey results. We present details on the McLEAP interface in Fig. S\ref{suppfig:quiz} (quiz question) and Fig. S\ref{suppfig:mcleap-interface} (order of problems and questions in an assignment). We finally show post-test results per sequence and content type first (Fig. S\ref{suppfig:post-test}), and per assignment and content type first (Fig. S\ref{suppfig:post-per-dataset}).

%More details on the intermediate McLeap results are shown in Figs. S\ref{suppfig:intermediate-test} and S\ref{suppfig:post-per-dataset}, which show test results per content order and dataset, and 

% \begin{figure}[!htb]
% \includegraphics[width=0.9\columnwidth]{figs/intermediate-test-per-content-order.pdf}
% \caption{Relative change from the mean grade of the intermediate tests grouped per content sequence and test (Data 2017-2018) showing a detailed breakdown of Fig. S\ref{suppfig:post-test} in the main text. Standard deviations are high, reflecting both variability in learning and the small number of questions per test (two in 2017, three in 2018).}
% \label{suppfig:intermediate-test}
% \end{figure}

